\documentclass[10pt,twocolumn,letterpaper]{article}

\usepackage[pagenumbers]{cvpr} 

\newcommand{\TODO}[1]{\textbf{\color{red}[TODO: #1]}}
\renewcommand{\TODO}[1]{}

\definecolor{cvprblue}{rgb}{0.21,0.49,0.74}
\usepackage[pagebackref,breaklinks,colorlinks,allcolors=cvprblue]{hyperref}

\usepackage{algorithm}
\usepackage{algpseudocode}
\usepackage[accsupp]{axessibility}

\def\paperID{} 
\def\confName{CVPR}
\def\confYear{2026}

\title{FluoCLIP: Stain-Aware Focus Quality Assessment in Fluorescence Microscopy}

\author{Hyejin Park$^{1}$\thanks{Equal contribution}\quad
Jiwon Yoon$^{1}$$^{*}$\quad
Sumin Park$^{1}$\quad
Suree Kim$^{3}$\quad
Sinae Jang$^{2}$ \quad
Eunsoo Lee$^{2}$\quad\\
Dongmin Kang$^{2}$$^{\dagger}$\quad
Dongbo Min$^{1}$\thanks{Corresponding Author}\\
\small$^{1}$Division of AI and Software, Ewha Womans University, South Korea\\  
\small$^{2}$Bioimaging Data Curation Center (BDCC), Department of Life Science, Ewha Womans University, South Korea\\
\small$^{3}$Analytical Solution Team, Daesang Corporation, South Korea\\
{\tt\small\{clrara, jwn, sm2023, jsa0801, eunssoo, dkang, dbmin\}@ewha.ac.kr},\quad {\tt\small kimsuree@hanmail.net}\\
}

\begin{document}
\maketitle
\begin{abstract}
Accurate focus quality assessment (FQA) in fluorescence microscopy is challenging due to stain-dependent optical variations that induce heterogeneous focus behavior across images. Existing methods, however, treat focus quality as a stain-agnostic problem, assuming a shared global ordering. 
We formulate \textbf{stain-aware FQA} for fluorescence microscopy, showing that focus-rank relationships vary substantially across stains due to stain-dependent imaging characteristics and invalidate this assumption. To support this formulation, we introduce \textbf{FluoMix}, the first dataset for stain-aware FQA spanning multiple tissues, fluorescent stains, and focus levels. 
We further propose \textbf{FluoCLIP}, a two-stage vision-language framework that grounds stain semantics and enables stain-conditioned ordinal reasoning for focus prediction, effectively decoupling stain representation from ordinal structure. By explicitly modeling stain-dependent focus behavior, FluoCLIP consistently outperforms both conventional FQA methods and recent vision-language baselines, demonstrating strong generalization across diverse fluorescence microscopy conditions. Code and dataset are publicly available at \url{https://fluoclip.github.io/}.
\end{abstract}

\section{Introduction}
\label{sec:intro}

Fluorescence microscopy enables high-resolution visualization of cellular and molecular structures by detecting light emitted from fluorophore-labeled targets, unlike bright-field imaging that relies on reflected or absorbed light~\cite{garcia2021novel, veronika2011correlation, mela2021application}. This capability makes it indispensable in biomedical research and digital pathology, where quantitative image analysis critically depends on acquiring in-focus images. However, automatic Focus Quality Assessment (FQA) in fluorescence microscopy remains largely underexplored.

\begin{figure}
\centering
\includegraphics[width=0.99\columnwidth]{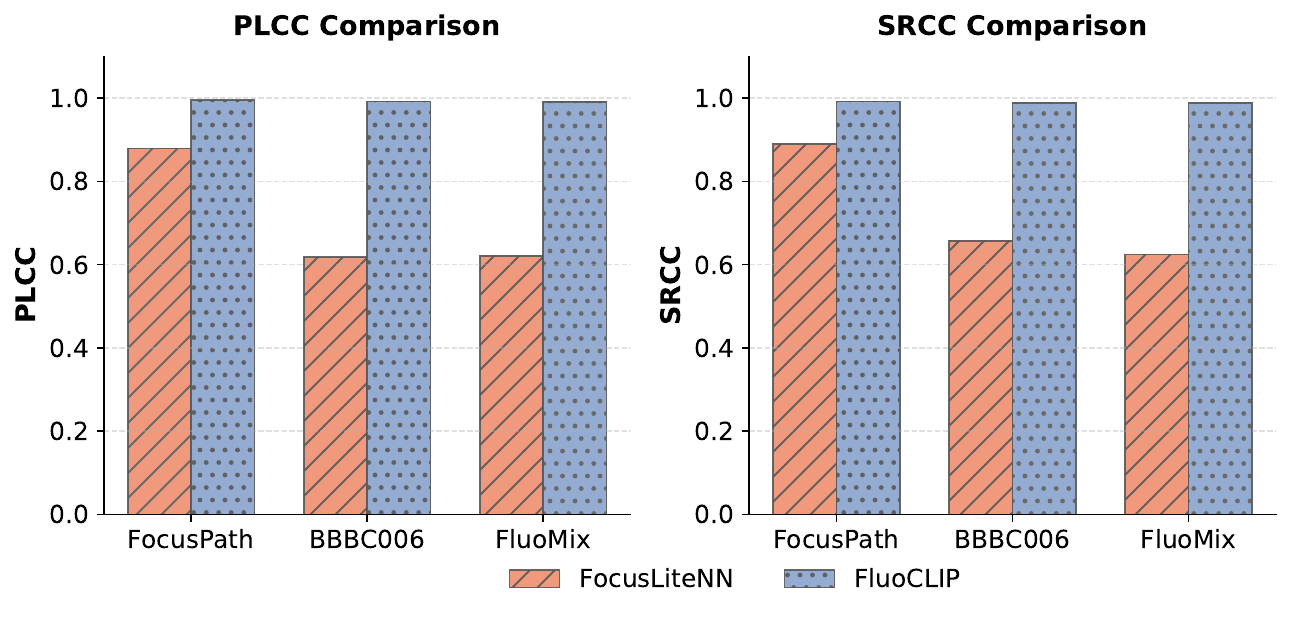}\\
\vspace{-1em}
\caption{\textbf{Modality gap in focus quality assessment.} Simple edge-based models such as FocusLiteNN~\cite{wang2020focuslitenn} perform reliably on bright-field datasets like FocusPath~\cite{hosseini2019encoding}, where blur is spatially uniform, but struggles on fluorescence datasets (BBBC006~\cite{ljosa2012annotated} and FluoMix) that exhibit stain-dependent, non-uniform defocus\textemdash underscoring the need for stain-aware FQA.}
\vspace{-1em}
\label{fig:focuslitenn}
\end{figure}


Focus degradation varies fundamentally between the two modalities. While bright-field microscopy blur is often spatially uniform, fluorescence microscopy exhibits stain-dependent and non-uniform blur due to differences in emission spectra, signal-to-noise ratio, and background fluorescence. This behavior arises from the wavelength-dependent diffraction limit ($d \propto \lambda / \mathrm{NA}$, where $\mathrm{NA}$ denotes the numerical aperture), leading to resolution differences across fluorophores with different emission wavelengths ($\lambda$). In our FluoMix ($\text{NA}=0.32$), fluorophores ranging from 405~nm (``DAPI") to 647~nm (``Alexa 647") exhibit a $\approx$1.6$\times$ variation in theoretical resolution. As a result, even perfectly focused images (Rank 0) inherently possess different high-frequency characteristics across stains, violating the underlying assumptions of stain-agnostic focus metrics. 

This modality gap poses critical challenges for FQA models. Edge-based approaches such as FocusLiteNN~\cite{wang2020focuslitenn}, which performs well on bright-field datasets like FocusPath~\cite{hosseini2019encoding}, fail to generalize to fluorescence datasets such as BBBC006~\cite{ljosa2012annotated} and FluoMix (Figure~\ref{fig:focuslitenn}). Their reliance on spatial gradients ignores stain-specific optical regimes, leading to unstable performance. These observations highlight the need for stain-aware modeling capable of handling heterogeneous focus degradation in fluorescence imaging.

Despite this need, existing datasets fail to capture the stain-dependent variability characteristic of fluorescence microscopy. FocusPath~\cite{hosseini2019encoding}, designed for bright-field imaging, assumes uniform optical conditions and lacks stain variation observed in fluorescence microscopy. BBBC006~\citep{ljosa2012annotated} contains homogeneous cell-line images acquired under tightly controlled in-vitro conditions where signal variation is minimal, offering limited biological and optical diversity. In contrast, tissue-level fluorescence microscopy involves complex three-dimensional structures, heterogeneous cell populations, and spatially varying stain distributions. These factors produce intricate light–matter interactions and non-uniform defocus patterns that reflect real challenges in biomedical imaging.

To comprehensively capture these factors, we construct \textbf{FluoMix}, a new dataset for Stain-Aware FQA in fluorescence microscopy. FluoMix contains multi-stain, tissue-level fluorescence images acquired under varying optical conditions, providing controlled yet diverse focus variations. By reflecting the spatial and biological heterogeneity of real tissue specimens, it establishes a practical foundation for developing and evaluating models capable of robust, stain-aware focus assessment.


Building upon this dataset, we introduce \textbf{FluoCLIP}, a vision-language framework for fluorescence FQA. Unlike bright-field imaging, focus perception depends not only on spatial sharpness but also on stain-dependent appearance, motivating semantic conditioning via text. However, pretrained CLIP~\citep{radford2021learning} lacks grounding in domain-specific stain concepts (e.g., ``DAPI" or ``Alexa-488"), and a single-stage alignment cannot disentangle stain semantics from focus-related variations. 

To address these limitations, FluoCLIP adopts a two-stage learning strategy. In the \textbf{stain-grounding phase}, learnable stain tokens are aligned with visual features to acquire fluorophore-specific semantics. In the subsequent \textbf{stain-guided ranking phase}, these learned stain embeddings guide ordinal focus prediction, enabling stain-conditioned focus modeling. 

We highlight our contributions as follows:
\begin{itemize}
\item We propose \textbf{FluoCLIP}, a two-stage ordinal vision–language framework that learns stain-specific grounding and stain-guided ranking for robust FQA.
\item We introduce \textbf{FluoMix}, a new dataset featuring diverse fluorescent stains and tissue-level focus variations, providing the first dataset for stain-aware FQA in fluorescence microscopy.
\item We formulate the task of \textbf{Stain-Aware FQA} in fluorescence microscopy, highlighting the need to model stain-dependent focus behavior. 
\end{itemize}

\section{Related Work}
\label{sec:related_work}
\subsection{Microscopy Focus Quality Assessment}

Early FQA methods adapted handcrafted metrics such as entropy or power spectrum statistics (e.g., PLLS~\citep{bray2012workflow}, statistical sharpness measures~\citep{koho2016image}) to detect focus loss, but lack generalizability across imaging protocols. Recent deep learning approaches improve robustness via ordinal regression~\citep{yang2018assessing}, specialized loss functions~\citep{albuquerque2022quasi}, label distribution learning~\citep{zhang2021diversity}, and sharp–blurred pair training~\citep{gu2023single}. However, these models are primarily optimized for bright-field imaging and remain stain-agnostic, failing to capture the stain-dependent focus behavior inherent in fluorescence microscopy. This gap necessitates modeling that integrates structural sharpness with stain-specific optical context.

\subsection{CLIP-based Ordinal Regression}

CLIP~\citep{radford2021learning} has been increasingly adapted for ordinal tasks by treating rank prediction as a vision–language matching problem. Prior works preserve ordinal structure through rank-aware prompting or interpolation, such as OrdinalCLIP~\citep{li2022ordinalclip} and L2RCLIP~\citep{wang2023learning}, while NumCLIP~\citep{du2024teach} improves numerical reasoning in score-based prediction tasks.  

Domain-specific variants such as BiomedCLIP~\citep{zhang2024biomedclip}, PubMedCLIP~\citep{eslami2023pubmedclip}, and PMC-CLIP~\citep{lin2023pmc} are pretrained CLIP models tailored for biomedical domains rather than ordinal regression methods. While they improve alignment for biomedical imagery, they are trained on semantic image–text pairs and do not explicitly model stain-dependent optical variations or ordinal focus structure required for FQA.

\section{Stain-Aware Focus Quality Assessment}
\label{sec:method}
\subsection{Problem Statement}
Focus Quality Assessment (FQA) in fluorescence microscopy can be naturally formulated as an ordinal regression task, where each $z$-stack slice is assigned a discrete focus level from a finite, ordered set. Let $\mathcal{X} = \{x_i\}_{i=1}^N$ denote a collection of fluorescence microscopy images, and $\mathcal{Y} = \{r_i\}_{i=1}^N$ the ground truth focus ranks, where $r_i \in \{1, 2, ..., K\}$ and $K$ denotes the total number of focus levels. The objective is to learn a model that maps each image $x_i$ to its ordinal rank $r_i$ by jointly capturing structural sharpness and stain-dependent semantics that influence focus perception.

Inspired by recent CLIP-based vision-language frameworks~\citep{radford2021learning}, we reformulate this problem as a text-image matching task. Specifically, we use a pretrained CLIP image encoder to extract visual features $v_i = \mathrm{E_{img}}(x_i)$, and construct a set of textual templates $\mathcal{P} = \{\mathcal{P}_1, \mathcal{P}_2, ..., \mathcal{P}_K\}$ that represent ordinal focus levels, such as ``fluorescence image with focus level 1" to ``fluorescence image with focus level $K$". 
These prompts are tokenized and encoded via the CLIP text encoder into classifier weights $t_k = \mathrm{E_{text}}(\mathcal{P}_k)$. The cosine similarity between image $x_i$ and rank prompt $\mathcal{P}_k$ is computed as:
\begin{equation}
z_{i,k} = \langle v_i, t_k \rangle, \quad \text{for}\;\;k=1,\ldots,K
\end{equation}
where $k$ indexes a focus level. The predicted focus level is obtained as
\begin{equation}
\hat{r}_i = \arg\max_k z_{i,k}.
\end{equation}


While this formulation enables CLIP to perform ordinal prediction, it assumes that both visual and textual embeddings are semantically aligned in a domain-agnostic manner. In fluorescence microscopy, however, focus quality is inherently stain-dependent due to wavelength dependent point spread function (PSF), leading to different degradation patterns across stains. As a result, a single global ordinal mapping becomes physically inconsistent. This motivates a stain-aware adaptation that grounds textual embeddings in fluorescence-specific semantics.

\subsection{FluoCLIP: Stain-Aware FQA Framework}

\begin{figure*}
\begin{minipage}[t!]{1\linewidth}
    \centering    \includegraphics[width=0.99\columnwidth]{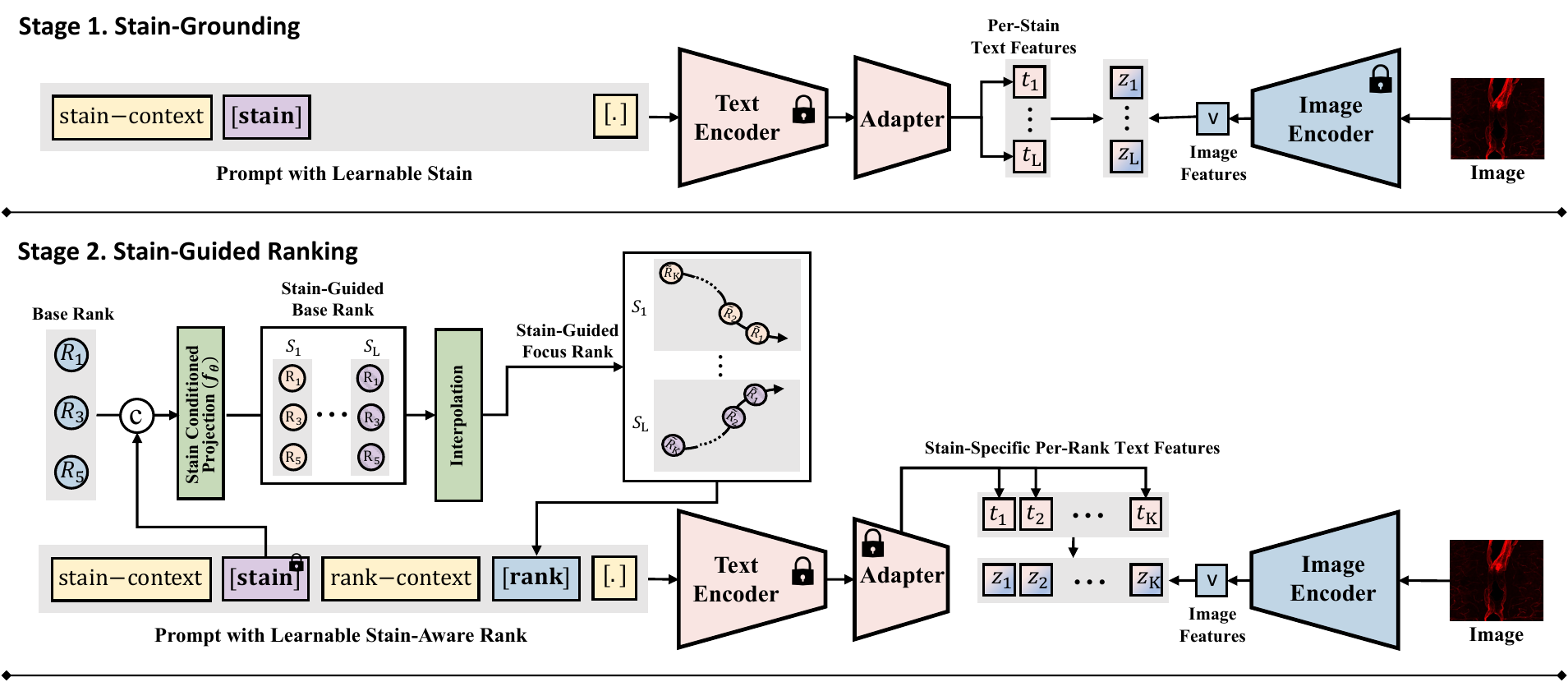}\\
    \vspace{-0.5em}
    \caption{{\bf Overview of FluoCLIP Framework.} FluoCLIP consists of two sequential stages for stain-aware focus quality assessment (FQA). In the \textit{Stage~1: Stain-Grounding}, learnable stain tokens are aligned with CLIP’s visual encoder through an adapter to form stain-specific textual features. In the \textit{Stage~2: Stain-Guided Ranking}, these grounded stain embeddings are used to condition rank prompts via interpolation and projection, producing stain-aware focus rank representations. Both stages jointly enable FluoCLIP to model stain-dependent focus behavior and align text–image features for stain-aware FQA. All parameters except explicitly noted as frozen were tuned in each stage.}
    \label{fig:fluoclip_overview}
\end{minipage}
\vspace{-0.5em}
\end{figure*}

We propose FluoCLIP, a two-stage vision-language framework for stain-aware focus quality assessment (Fig.~\ref{fig:fluoclip_overview}). In the first stage, the model performs \textbf{stain grounding}, learning to align textual stain tokens with CLIP visual representations so that the text encoder acquires fluorescence-specific semantics. Rather than fine-tuning the CLIP text encoder, which may cause semantic drift, we keep it frozen and attach a compact adapter that learns stain-specific attributes. This design preserves the linguistic consistency of the pretrained text encoder while enabling controlled adaptation with diverse stains. In the second stage, \textbf{stain-guided ranking}, the learned stain embeddings are used to condition focus prediction on stain-dependent appearance variations. Through this process, FluoCLIP learns how focus perception varies not only with the degree of defocus but also with the emission and scattering characteristics unique to each fluorophore.

\subsubsection{Stage 1: Stain-Grounding}


The pretrained CLIP text encoder, trained on general web data with no exposure to fluorescence-related terminology or visual semantics, is inherently ill-suited for modeling the fluorescence-specific characteristics of stains. As shown in Table~\ref{tab:ablation}, directly inserting stain tokens into textual prompts without adaptation results in a noticeable drop in performance, confirming this domain gap. To address this, we introduce a stain-grounding stage that explicitly aligns textual stain embeddings with CLIP visual encoder. Instead of fine-tuning the pretrained text encoder, which often causes semantic drift and degrades the stability of multimodal alignment, we freeze the pretrained text encoder and instead optimize newly introduced stain embeddings and lightweight adapter layers appended to the text encoder.

Given a predefined set of $L$ stains, each stain description is tokenized and embedded using CLIP’s token embedding layer. Let $\mathbf{S} = \{\mathbf{S}_{l}\}^L_{l=1}$ denote a collection of learnable stain embeddings, where each $S_l \in \mathbb{R}^{C\times D}$ represents the token-level embedding sequence for stain $s_l$ ($C$: the number of tokens per stain, $D$: embedding dimension).
These embeddings are grounded to the CLIP visual encoder to encode fluorophore-specific characteristics. 

To integrate stain tokens into CLIP’s textual space, we construct pseudo-sentences of the form (Fig.~\ref{fig:fluoclip_overview}):
\begin{equation}
\mathcal{P}^{\text{stain}}_l = \texttt{Concat}\big([\textit{context}], [\mathbf{S}_l]\big), \quad \text{for}\;\;l=1,...,L,
\label{eq:prompt_stain}
\end{equation}
where \([\textit{context}]\) denotes the base prompt and $[\mathbf{S}_l]$ the learnable stain tokens. The operator $\texttt{Concat}(\cdot)$ represents token-level concatenation along the sequence dimension. 

A lightweight adapter module, consisting of a single self-attention layer and a two-layer MLP, facilitates interaction between context and stain tokens, enabling the encoder to refine their joint representation. By optimizing only these stain-related parameters, the model learns stain-aware textual embeddings aligned with CLIP visual features, providing fluorescence-specific semantic grounding and a biologically meaningful initialization for the subsequent stain-guided ranking stage. 

Given an image $x_i$ with the ground truth stain label $s_i$, its visual feature is obtained as 
$v_i = \mathrm{E_{img}}(x_i)$.
From the text side, the stain prompts 
$\mathcal{P}^{\text{stain}} = \{\mathcal{P}^{\text{stain}}_1, \ldots, \mathcal{P}^{\text{stain}}_L\}$ 
in \eqref{eq:prompt_stain} are encoded by the text encoder as 
$T^{\text{stain}} = \mathrm{E_{text}}(\mathcal{P}^{\text{stain}})$.
The cosine similarity between $x_i$ and each stain prompt is then computed as
\begin{equation}
z_{i,l}^{\text{stain}} = \langle v_i, t_l^{\text{stain}} \rangle, \quad \text{for}\;\;l=1,...,L,
\label{eq:stain_sim}
\end{equation}
where $t_l^{\text{stain}} \in T^{\text{stain}}$. 
The alignment is optimized in a way of predicting the stain label using the cross-entropy and KL-divergence objectives described in Section~\ref{sec:training_objective}.
This process aligns each image feature with the correct stain embedding, encouraging the model to capture fluorophore-specific semantics in the shared vision–language space. 

\subsubsection{Stage 2: Stain-Guided Ranking}
In fluorescence microscopy, each stain exhibits a distinct relationship between focus and image appearance, producing stain-specific feature distributions that a single shared ranking space cannot capture. To handle this heterogeneity, we introduce a stain-guided embedding space that conditions the rank representation on the stain identity learned in Stage 1. 

To preserve the ordinal continuity between focus levels, we follow the interpolation strategy introduced in OrdinalCLIP~\cite{li2022ordinalclip}. A small number of base rank embeddings represent boundary ranks (\emph{e.g.}, most- and least-focused), and intermediate ranks are obtained through weighted interpolation between them. This design encourages smooth transitions in the rank embedding space and stabilizes ordinal reasoning across focus levels.


Let $\mathbf{R}^{\text{base}}=\{\mathbf{R}_{k'}^{\text{base}}\}_{{k'}=1}^{K'}$ denote $K'$ base rank embeddings, where $\mathbf{R}_{k'}^{\text{base}} \in \mathbb{R}^{C\times D}$. Given the stain embedding $\mathbf{S}_l$ learned in Stage 1 for $l=1,...,L$, a lightweight conditioning network $f_\theta$ projects each base rank embedding into a stain-guided space:
\begin{equation}
\vspace{-0.5em}
\mathbf{R}_{k'}^{l} = f_\theta(\mathbf{R}_{k'}^{\text{base}}, \, \mathbf{S}_l).
\label{eq:2nd_feat}
\end{equation}
The resulting $\mathbf{R}_{k'}^{l}$ represents the stain-conditioned rank embedding. Each base rank prompt is modulated by the $l^{th}$ stain semantics, enabling FluoCLIP to capture stain-specific focus behavior.

Finally, within each stain we generate intermediate ranks by interpolating the stain-guided base ranks:
\begin{equation} \label{eq:rank_tilde}
\tilde{\mathbf{R}}_{k}^{l}=\sum_{{k'}=1}^{K'}w_{r,{k'}}\,\mathbf{R}_{k'}^{l}, \quad \text{for}\;\; k=1,...,K,
\end{equation}
where $w_{r,{k'}}$ is a predefined interpolation weight for an inverse interpolation between base ranks~\cite{li2022ordinalclip}.


For an input image $x_i$ with stain label $s_i$, the stain-conditioned rank embeddings are inserted into pseudo-sentence templates (Figure~\ref{fig:fluoclip_overview}):
\begin{equation} \label{eq:prompt_rank}
\mathcal{P}_{k}^{\text{rank},i}
=\texttt{Concat}\big([\textit{stain-context}],
[\mathbf{S}_{s_i}],
[\textit{rank-context}],
[\tilde{\mathbf{R}}_{k}^{s_i}]\big),
\end{equation} 
where $\mathbf{S}_{s_i}$ is the stain embedding corresponding to the stain $s_i$ of image $x_i$, 
and $\tilde{\mathbf{R}}_{k}^{s_i}$ is the $k$-th interpolated rank embedding conditioned on stain $s_i$. 
The resulting prompt sequence $\mathcal{P}_{k}^{\text{rank},i}$ is then passed through the CLIP text encoder to obtain the final rank-aware textual representation, which is inherently sample-dependent due to its reliance on the stain $s_i$.
A set of stain-guided rank prompts 
$\mathcal{P}^{\text{rank},i}=\{\mathcal{P}_{1}^{\text{rank},i},\ldots,\mathcal{P}_{K}^{\text{rank},i}\}$ 
is encoded as 
\begin{equation}
t^{\text{rank},i}=\mathrm{E_{text}}(\mathcal{P}^{\text{rank},i}),
\end{equation}
where $t^{\text{rank},i} = \{t_1^{\text{rank},i},...,t_K^{\text{rank},i} \}$.
The similarity between $x_i$ and each rank prompt is defined as
\begin{equation}
z_{k}^{\text{rank},i}=\langle v_i,t_{k}^{\text{rank},i}\rangle, \quad \text{for}\;\; k=1,...,K,
\label{eq:rank_sim}
\end{equation}
where $\langle\cdot,\cdot\rangle$ denotes cosine similarity, and $v_i=\mathrm{E_{img}}(x_i)$ is the visual embedding. 
These similarities are trained to predict the focus label using the cross-entropy and KL-divergence objectives described in Section~\ref{sec:training_objective}, with the ground truth focus label $r_i \in\{1,...,K\}$.

\begin{figure*}[t!]
  \centering
  \begin{subfigure}[t]{0.24\textwidth}
    \centering
    \includegraphics[width=\textwidth]{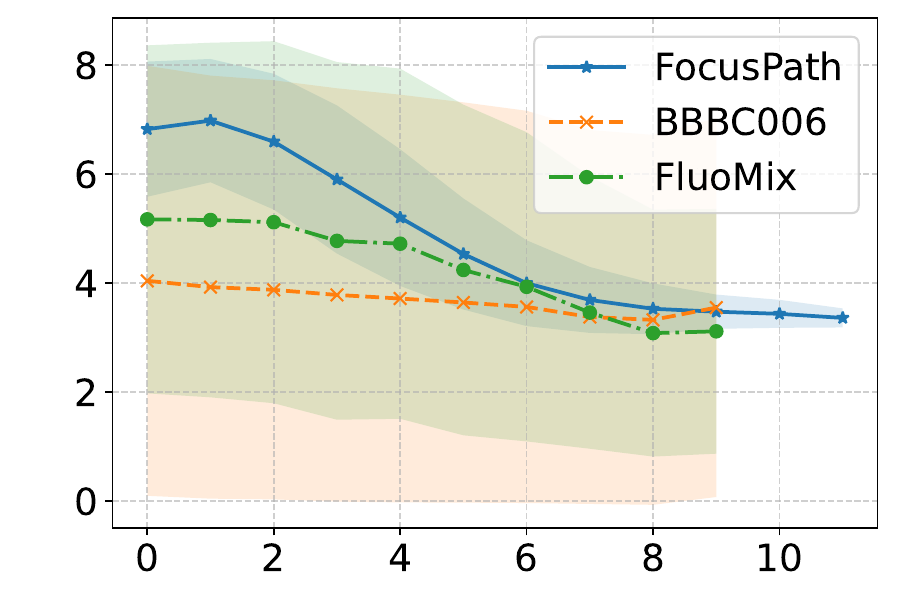}
    \caption{Rank Curve}
    \label{fig:rank_curve}
  \end{subfigure}
  \centering
  \begin{subfigure}[t]{0.24\textwidth}
    \centering
    \includegraphics[width=\textwidth]{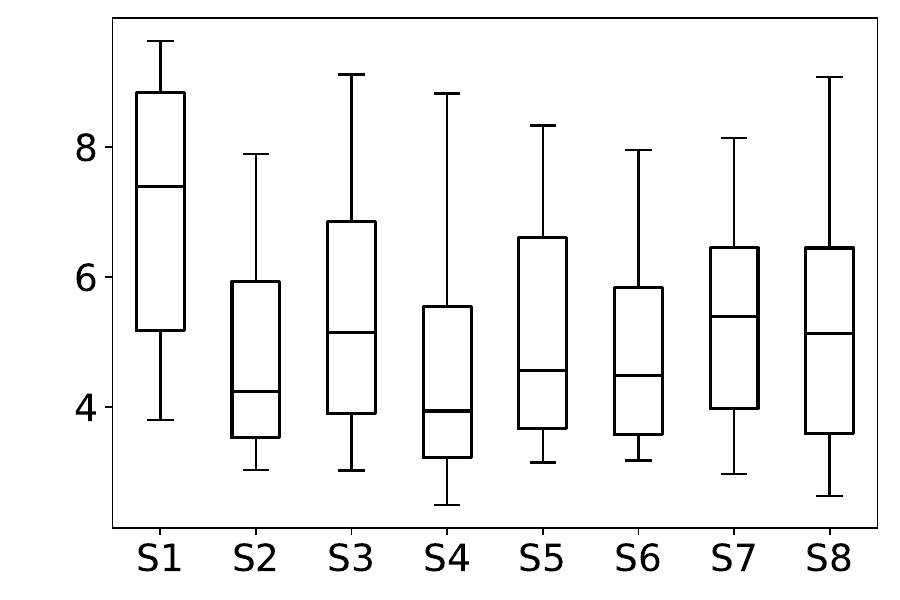}
    \caption{FocusPath}
    \label{fig:focuspath}
  \end{subfigure}
  \begin{subfigure}[t]{0.24\textwidth}
    \centering
    \includegraphics[width=\textwidth]{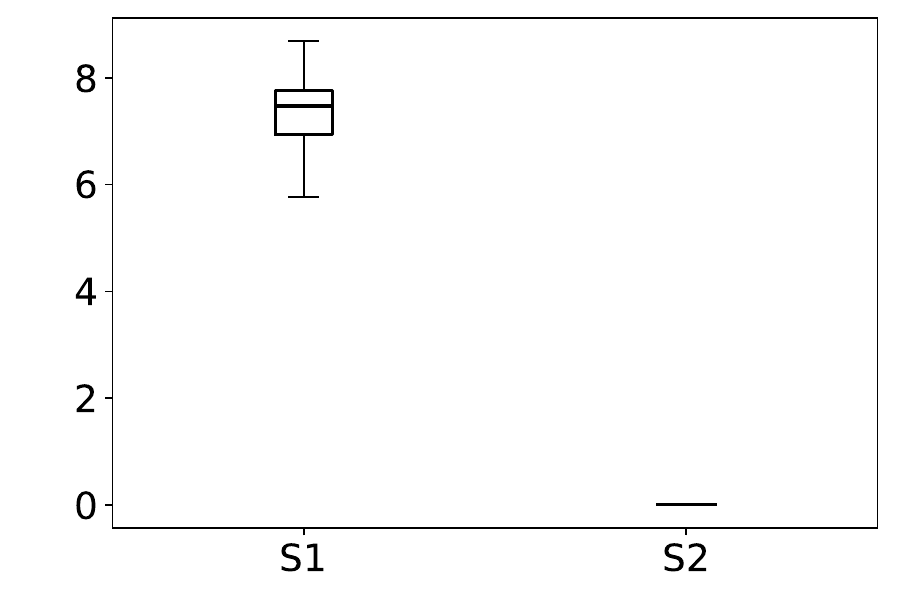}
    \caption{BBBC006}
    \label{fig:bbbc}
  \end{subfigure}
  \begin{subfigure}[t]{0.24\textwidth}
    \centering
    \includegraphics[width=\textwidth]{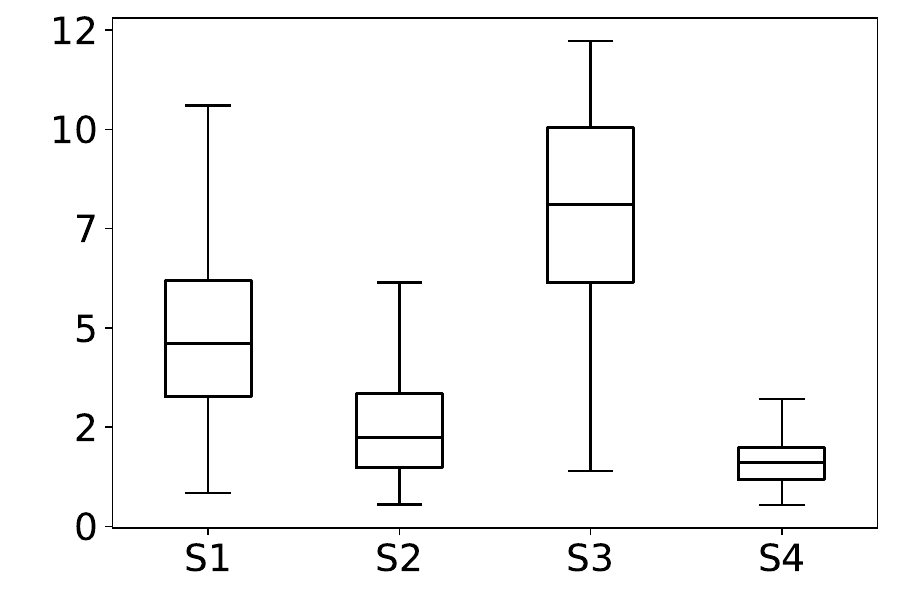}
    \caption{FluoMix}
    \label{fig:fluomix}
  \end{subfigure}
  \vspace{-0.8em}
  \caption{Empirical Analysis of Stain-Dependent Focus Behavior: (a) Mean spatial frequency (SF) versus focus rank for three datasets; the shaded region indicates $\pm$1 standard deviation across samples. SF decreases monotonically with increasing rank, confirming that SF reliably captures focus degradation.
  (b)--(d) Boxplots of SF values across stains for each dataset (x-axis: stain identity, y-axis: SF distribution). FocusPath shows stain-invariant SF trends, wherease BBBC006 and FluoMix display pronounced stain-dependent variability.}
  \label{fig:boxplot}
  \vspace{-0.5em}
\end{figure*}

\subsection{Overall Training Objective}
\label{sec:training_objective}

\noindent\textbf{Loss Functions.} 
FluoCLIP produces two types of similarities during training: 
stain-alignment similarities $z_{i,l}^{\text{stain}}$ from Stage~1 (Eq.~\ref{eq:stain_sim}) and stain-guided ranking similarities $z_{i,r}^{\text{rank}}$ from Stage~2 (Eq.~\ref{eq:rank_sim}). 
Following \citep{li2022ordinalclip}, both are optimized using two complementary objectives: a cross-entropy loss $\mathcal{L}_{CE}$ for image-to-text alignment and  
a KL-divergence loss $\mathcal{L}_{KL}$ to enforce ordinal consistency in the similarity distributions.
The full training objective combines the two losses:
\begin{equation}
\mathcal{L}_{\text{total}}
= \alpha\cdot \mathcal{L}_{CE}
+ \beta\cdot \mathcal{L}_{KL},
\end{equation}
where $\alpha$ and $\beta$ are set to $1.0$ in our experiments.
\section{FluoMix for Stain-Aware FQA}
\label{sec:dataset}

\subsection{Motivation and Design Principles}
Existing datasets for focus quality assessment (FQA) are limited in scope and fail to represent the diversity of fluorescence microscopy. FocusPath~\cite{hosseini2019encoding} is a bright-field histopathology dataset containing nine differently stained human tissues, each with 14 z-stack slices. Although it provides multiple stains, focus transitions are smooth and staining conditions are consistent across slides, making the dataset less challenging for models that rely primarily on low-level contrast cues. BBBC006~\cite{ljosa2012annotated}, on the other hand, is a fluorescence microscopy dataset of U2OS cancer cells stained with Hoechst 33342 and phalloidin. While it captures true fluorescence imaging conditions, it includes only two stains and homogeneous in-vitro cell structures, limiting its biological and optical diversity.  

Consequently, existing datasets do not reflect the complexity of real-world fluorescence imaging, where multiple tissues, stains, and acquisition protocols coexist. To fill this gap, we introduce FluoMix\textemdash a multi-tissue, multi-stain dataset specifically designed for stain-aware FQA in fluorescence microscopy.

\begin{figure*}
\begin{minipage}[t!]{1\linewidth}
    \centering    
    \includegraphics[width=0.95\columnwidth]{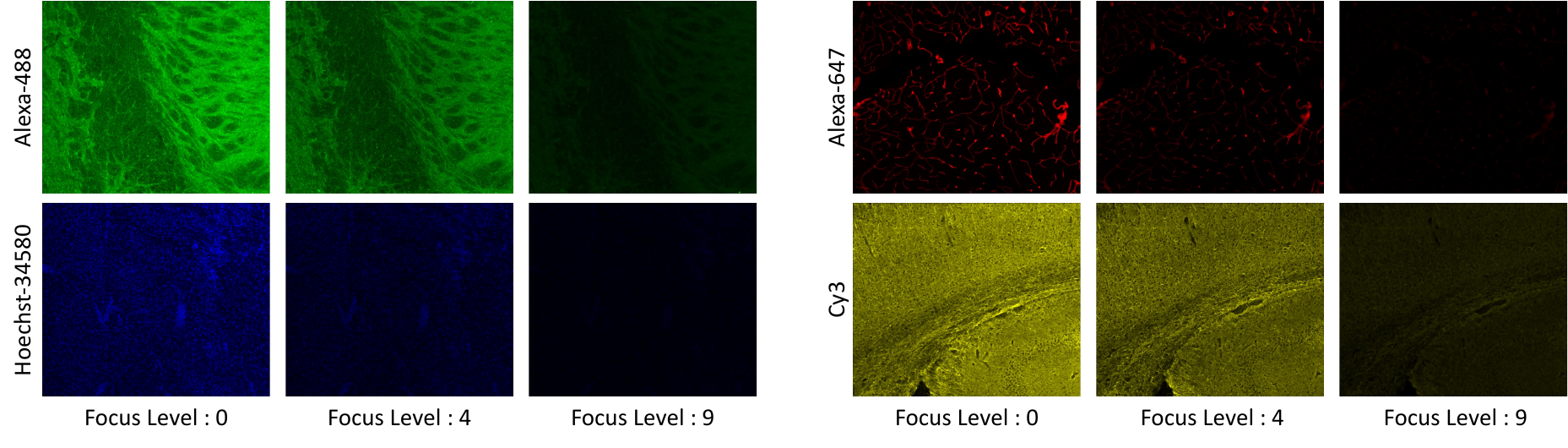}\\
    \vspace{-0.5em}
    \caption{\textbf{Stain-dependent appearance across focus levels.} Images from different stains exhibit distinct visual characteristics across focus levels due to wavelength-dependent optical effects ($d \propto \lambda / \text{NA}$). Focus level 0 corresponds to well-focused images, while higher levels (e.g., level 9) indicate increasing defocus. These differences indicate that identical focus levels can correspond to different visual patterns across stains, challenging the assumption of a universal ordinal mapping.}
    \label{fig:fluomix_evidence}
\end{minipage}
\vspace{-0.5em}
\end{figure*}

\subsection{FluoMix Composition}
FluoMix aggregates fluorescence microscopy images across multiple tissue types and staining conditions to capture diverse optical and biological characteristics. 
Specifically, it includes images from brain, lung, and liver tissues, with up to four distinct stains per sample. 
Each field of view (FOV) is acquired as a complete 34-plane z-stack, covering full range from sharp to severely blurred slices. 
In total, FluoMix contains approximately 56,000 images spanning 28 distinct imaging channels defined by stain–target–tissue combinations. These include 3 tissue types, 6 fluorophores, and 13 biological targets covering diverse functional categories such as neuronal, vascular, microglial, and epithelial markers. This diversity introduces heterogeneous optical regimes, making focus behavior inherently stain-dependent. This composition results in a broad range of optical and biological variability, reflecting diverse fluorescence imaging conditions (see Table~\ref{tab:fluomix_overview} and Section~\ref{sec:fluomix_details}).

This design allows FluoMix to explicitly model \textit{stain-dependent focus degradation}, providing a realistic dataset for testing Stain-Aware FQA methods. Alongside the raw images, we provide detailed \textbf{metadata} describing the stain identity, target protein, and acquisition protocol, enabling multimodal research, including vision–language modeling, beyond conventional image-only baselines. 
Representative visual examples are provided in supplementary material.

\subsection{Annotation Protocol}
We adopt a relative annotation scheme inspired by prior work~\citep{hosseini2019encoding}. 
For each z-stack, the visually best-focused slice is selected by an expert as a reference, and all other slices are assigned \textbf{relative focus levels} from $0$ (best focus) to $R{-}1$ (worst). In this release, we provide annotations with $R=10$ discrete ordinal levels derived from the relative position within the z-stack. 

This strategy offers a consistent yet scalable protocol for large-scale datasets, while maintaining ordinal relationships essential for FQA. Importantly, the full raw z-stacks are released so that future work can explore alternative annotation schemes such as continuous scores, perceptual metrics, or data condensation approaches-without re-acquisition overhead. By decoupling dataset release from a fixed labeling protocol, FluoMix provides a flexible foundation for future research in \emph{stain-aware FQA modeling}.

\section{Empirical Evidence for Stain-Dependent Focus Behavior}
\label{sec:analysis}

Having introduced the dataset design, we next analyze how focus--rank relationships vary across FocusPath~\cite{hosseini2019encoding}, BBBC006~\cite{ljosa2012annotated}, and our proposed FluoMix to empirically validate the need for stain-aware focus modeling. 
The spatial frequency (SF) metric~\cite{Eskicioglu1995ImageQM} quantifies the density of high-frequency components in an image. Since well-focused images preserve high-frequency details while defocus attenuates them, SF serves as an effective proxy for sharpness and focus quality. We therefore use the spatial frequency (SF) metric as a quantitative measure to compare how focus behavior differs between bright-field and fluorescence modalities. Since the point spread function (PSF) determines the spatial frequency response of an imaging system, wavelength-dependent PSF directly affects the high-frequency content captured by the SF metric. Further explanation of SF metric are provided in the supplementary material.

\vspace{-0.4em}
\subsection{Per-Rank Behavior of Sharpness Metric}
We first examine the relationship between the spatial frequency (SF) metric and the discrete focus ranks to verify that the metric reliably reflects the degree of defocus represented by the ranks. As shown in Figure~\ref{fig:boxplot}\subref{fig:rank_curve}, the mean SF values exhibit a generally monotonic decrease as the focus level moves from in-focus to defocused regions across all three datasets. This confirms that the SF metric can reasonably capture the overall directionality of focus degradation implied by the ordinal labels, validating its use as a quantitative sharpness indicator for subsequent analysis. In other words, while focus ranks are discrete and manually defined, the SF metric provides an objective, continuous indicator of focus quality, exhibiting a strong correlation with the discrete rank progression.

\subsection{Stain-Wise Variability Across Datasets}
We then analyze the distribution of the SF metric across different stains within each dataset in Figure~\ref{fig:boxplot}. 
In FocusPath~\cite{hosseini2019encoding}, which consists of bright-field H\&E-stained whole-slide images, the per-stain variance is modest, with stains exhibiting closely aligned SF ranges and largely overlapping distributions, indicating stain-invariant focus behavior.
In contrast, BBBC006~\cite{ljosa2012annotated} and FluoMix exhibit substantial stain-dependent variability. Different fluorescent stains occupy distinct SF ranges with largely non-overlapping variances, demonstrating that focus degradation is highly stain-dependent in fluorescence microscopy. This observation supports the hypothesis that focus quality should be modeled as a function of stain characteristics rather than assuming a universal focus-rank mapping. 

\subsection{Sharpness Metric and Rank Correlation}
To further quantify this effect, we compute the Spearman rank correlation coefficient (SRCC) between focus rank and SF metric for each stain. Since spatial frequency decreases as defocus increases, SF monotonically decreases with rank. This produces an inherently negative SRCC value, where the absolute magnitude $|\rho|$ reflects the strength of the relationship: larger values indicate that SF aligns more consistently with the underlying focus levels. 

As shown in Table~\ref{tab:sf_spearman}, FocusPath~\cite{hosseini2019encoding} achieves the highest correlation, indicating that its ordinal labels correspond closely to actual blur magnitudes under stain-invariant conditions. However, correlations drop significantly in BBBC006~\cite{ljosa2012annotated} and FluoMix, revealing that rank–blur consistency weakens in fluorescence microscopy due to heterogeneous optical properties across stains. This reduced correlation implies that a single, stain-agnostic ranking model cannot reliably capture focus variation across fluorescence conditions. 

This inconsistency arises because identical focus levels can produce different spatial frequency profiles across stains, causing a global ranking function to misinterpret stain-dependent sharpness variations. Fig.~\ref{fig:fluomix_evidence} provides visual evidence that images at the same ordinal focus level exhibit distinct appearance patterns across stains. These findings highlight the necessity of stain-aware modeling for reliable focus assessment in fluorescence microscopy.

\begin{table}[t]
\centering
\caption{Spearman Rank Correlation Coefficient (SRCC) statistics of the Spatial Frequency (SF) metric and focus rank. The results reported are mean $\pm$ standard deviation across all stains in each datasets.}
\vspace{-0.8em}
\scriptsize
\addtolength{\tabcolsep}{-1pt}
\begin{tabular}{lccc}
\toprule
\textbf{Metric} & \textbf{FluoMix} & \textbf{FocusPath} & \textbf{BBBC006} \\
\midrule
SRCC $\rho$ & $-0.528\pm0.0940$ & $-0.840\pm0.0923$ & $-0.343\pm0.2920$ \\
\bottomrule
\end{tabular}
\label{tab:sf_spearman}
\vspace{-0.5em}
\end{table}


\begin{table}[t!]
\centering
    \caption{Comparison of FluoCLIP with existing methods using a ResNet50 image encoder on FluoMix. The results are mean $\pm$ standard deviation over five runs with different random seeds.}
    \vspace{-0.8em}
    \scriptsize\addtolength{\tabcolsep}{-2.5pt}
    \begin{tabular}{lcccc}
    \toprule
    Method  &  Accuracy (\%) & PLCC $\uparrow$ & SRCC $\uparrow$ & MAE $\downarrow$ \\ 
    \midrule
    FocusLiteNN & - & 0.621$\pm$0.0042 & 0.624$\pm$0.0047 & 1.610$\pm$0.0058 \\
    \midrule
    CE & 54.59$\pm$1.42 & 0.952$\pm$0.0023 & 0.957$\pm$0.0017 & 0.510$\pm$0.0148  \\
    OE & 63.46$\pm$1.38 & 0.966$\pm$0.0019 & 0.969$\pm$0.0017 & 0.395$\pm$0.0171 \\
    CO2 & 47.26$\pm$1.99 & 0.927$\pm$0.0054 & 0.931$\pm$0.0052 & 0.653$\pm$0.0335 \\
    QULCE & 55.55$\pm$2.12 & 0.951$\pm$0.0050 & 0.957$\pm$0.0041 & 0.512$\pm$0.0320 \\
    \midrule
    CLIP & 81.59$\pm$0.52 & 0.988$\pm$0.0007  & 0.987$\pm$0.0007 & 0.187$\pm$0.0056 \\
    OrdinalCLIP & 83.12$\pm$0.41 & 0.989$\pm$0.0006 & 0.988$\pm$0.0005 & 0.172$\pm$0.0049\\
    NumCLIP & - & 0.985$\pm$0.0004 & 0.985$\pm$0.0005 & 0.325$\pm$0.0054  \\
    FluoCLIP    &\textbf{85.21$\pm$0.88} & \textbf{0.991$\pm$0.0006} & \textbf{0.989$\pm$0.0004} & \textbf{0.150$\pm$0.0087}\\
    \bottomrule
    \end{tabular}
    \label{tab:fluomix}
\vspace{-0.8em}
\end{table}

\begin{table}[t!]
\centering
\caption{Performance comparison of CLIP-based models with existing methods using a ViT/B-16 image encoder on FluoMix. The results are mean $\pm$ standard deviation over five runs with different random seeds.}
\vspace{-0.8em}
\scriptsize\addtolength{\tabcolsep}{-2pt}
\begin{tabular}{lcccc}
\toprule
Method  &  Accuracy (\%)  & PLCC $\uparrow$ & SRCC $\uparrow$ & MAE $\downarrow$ \\
\midrule 
CLIP & 89.13$\pm$0.26 & 0.993$\pm$0.0002 & 0.991$\pm$0.0001 & 0.110$\pm$0.0029  \\   
OrdinalCLIP & 89.95$\pm$0.31 & 0.993$\pm$0.0002 & 0.991$\pm$0.0002 & 0.102$\pm$0.0032  \\ 
NumCLIP & - & 0.987$\pm$0.0014 & 0.988$\pm$0.0003 & 0.133$\pm$0.0031  \\    
BiomedCLIP & 83.99$\pm$0.25 & 0.990$\pm$0.0003 & 0.988$\pm$0.0003 & 0.163$\pm$0.0027 \\    
FluoCLIP  & \textbf{90.75$\pm$0.52} & \textbf{0.994$\pm$0.0005} & \textbf{0.992$\pm$0.0004} & \textbf{0.093$\pm$0.0055} \\ 
\bottomrule
\end{tabular}
\label{tab:fluomix_vit}
\vspace{-0.8em}
\end{table}

\section{Experiments}
\label{sec:experiments}

\subsection{Datasets and Experiment Settings}

\noindent\textbf{Datasets.} All experiments were conducted with consistent training protocols across FocusPath~\citep{hosseini2019encoding}, BBBC006~\citep{ljosa2012annotated}, and FluoMix. 
Within FluoMix, brain tissue images were used for training, and images from other tissues were used for validation and few-shot testing to assess cross-tissue generalization. More details are provided in the supplementary material.

\noindent\textbf{Training Details.} We adopted CLIP~\cite{radford2021learning} as backbone, using either CLIP-pretrained ResNet50~\citep{he2016deep} or ViT-B/16~\citep{dosovitskiy2020image} image encoders with pretrained weights. FluoCLIP is trained with a two-stage training protocol consisting of 10 epochs for the \textit{Stain-Grounding} and 100 epochs for the \textit{Stain-Guided Ranking}. All models are trained under identical settings for fair comparison, including data preprocessing and augmentations. Input images were resized to $256\times256$, randomly cropped to $224\times224$, and augmented using horizontal flips and standard normalization.

\noindent\textbf{Baselines.} We compared our FluoCLIP against FocusLiteNN~\citep{wang2020focuslitenn}, OrdinalCLIP~\citep{li2022ordinalclip}, NumCLIP~\citep{du2024teach}, BiomedCLIP~\cite{zhang2024biomedclip} and standard baselines including Cross-Entropy (CE), Ordinal Encoding (OE)~\citep{cheng2008neural}, CO2~\citep{albuquerque2021ordinal}, and QULCE~\citep{albuquerque2022quasi}. For fair comparison, all baselines were experimented using the same ResNet50 encoder and respective loss functions. FocusLiteNN was evaluated separately using its original lightweight CNN architecture due to structural incompatibility with our framework.

\noindent\textbf{Evaluation Metrics.} Performance was evaluated using classification accuracy, PLCC (Pearson linear correlation), SRCC (Spearman rank correlation), and MAE (mean absolute error). To compute PLCC, SRCC, and MAE, the predicted focus level is given by $\hat{y} = \sum_{j=0}^{M-1} p_j \cdot j$, where $p_j$ is the probability of focus level $j$, and $M$ is the total number of discrete ranks.

\begin{table}[t!]
\centering
    \caption{Comparison of FluoCLIP with existing methods using a ResNet50 image encoder on BBBC006~\cite{ljosa2012annotated}. The results are mean $\pm$ standard deviation over five runs with different random seeds.}
    \vspace{-0.8em}
    \scriptsize\addtolength{\tabcolsep}{-2pt}
    \begin{tabular}{lcccc}
    \toprule
    Method  &  Accuracy (\%) & PLCC $\uparrow$ & SRCC $\uparrow$ & MAE $\downarrow$ \\
    \midrule
    FocusLiteNN & - &  0.618$\pm$0.0062 & 0.657$\pm$0.0080 & 1.599$\pm$0.6571 \\
    \midrule
    CE & 75.71$\pm$8.24 & 0.973$\pm$0.0102 & 0.975$\pm$0.0091 & 0.266$\pm$0.0928 \\
    OE & 75.03$\pm$12.55 & 0.967$\pm$0.0176 & 0.969$\pm$0.0159 & 0.296$\pm$0.1565 \\
    CO2 & 68.84$\pm$3.91 & 0.965$\pm$0.0056 & 0.968$\pm$0.0050 & 0.347$\pm$0.0469 \\
    QULCE & 75.25$\pm$7.57 & 0.962$\pm$0.0123 & 0.965$\pm$0.0109 & 0.301$\pm$0.1069  \\
    \midrule
    OrdinalCLIP & 90.67$\pm$0.54 & 0.991$\pm$0.0004 & \textbf{0.989$\pm$0.0002} & 0.104$\pm$0.0058\\
    NumCLIP & - & 0.973$\pm$0.0009 & 0.980$\pm$0.0006 & 0.249$\pm$0.0051 \\   
    FluoCLIP    &\textbf{93.05$\pm$0.34} & \textbf{0.992$\pm$0.0005} & \textbf{0.989$\pm$0.0004} & \textbf{0.080$\pm$0.0046}\\ 
    \bottomrule
    \end{tabular}
    \label{tab:bbbc}
\end{table}

\begin{table}[t!]
\centering
    \caption{Comparison of FluoCLIP with existing methods using a ResNet50 image encoder on FocusPath~\cite{hosseini2019encoding}. The results are mean $\pm$ standard deviation over five runs with different random seeds.}
    \vspace{-0.8em}
    \scriptsize\addtolength{\tabcolsep}{-2pt}
    \begin{tabular}{lcccc}
    \toprule
    Method  &  Accuracy (\%) & PLCC $\uparrow$ & SRCC $\uparrow$ & MAE $\downarrow$ \\
    \midrule
    OrdinalCLIP & 94.98$\pm$0.99 & 0.997$\pm$0.0006 & 0.993$\pm$0.0004 & 0.051$\pm$0.0099  \\
    FluoCLIP    & 91.11$\pm$1.40 & 0.995$\pm$0.0007 & 0.992$\pm$0.0005 & 0.092$\pm$0.0142 \\ 
    \bottomrule
    \end{tabular}
    \label{tab:focuspath}
    \vspace{-0.5em}
\end{table}

\begin{table*}[t!]
\centering
\caption{Ablation study on FluoCLIP components on the FluoMix dataset. The baseline (OrdinalCLIP~\citep{li2022ordinalclip}) includes only the focus token. We incrementally add grounded stain tokens (C) and stain-guided rank modules (D) to evaluate their individual contributions.}
\vspace{-0.8em}
\scriptsize\addtolength{\tabcolsep}{4pt}
\begin{tabular}{c|l|ccc|cc|c|c|c}
\toprule
Type & Configuration &$S^{\text{plain}}$ & $S^{\text{train}}$ & $S$ & $R$ & $\tilde{R}^S$  & Acc. (\%) & Step Gain  & Total Gain\\
\midrule
(A) & Baseline (OrdinalCLIP) with Rank ($R$)        &&&                 & \checkmark &         & 83.12$\pm$0.41 & - & -\\
(B) & (A) + Plain Stain Token ($S^{\text{plain}}$)& \checkmark &&       & \checkmark &         & 83.21$\pm$3.93&0.09 &-\\
(C) & (A) + Learnable Stain Token ($S^{\text{train}}$)&& \checkmark&    & \checkmark &         & 81.38$\pm$0.63&-1.74 &-\\
(D) & (A) + Grounded Stain Token ($S$)          &&& \checkmark          & \checkmark &         & 84.28$\pm$0.88&1.16&1.16\\
(E) & (D) + Stain-Guided Rank Token ($\tilde{R}^S$)& & & \checkmark     && \checkmark          & 85.21$\pm$0.88&0.93 &2.09\\
\bottomrule
\end{tabular}
\label{tab:ablation}
\vspace{-0.5em}
\end{table*}

\subsection{Comparative Study}

\textbf{Results on FluoMix.} FluoMix presents the greatest challenge due to its high stain variability and heterogeneous tissue structures. Traditional models exhibit significant drops in performance, with accuracy ranging from 47.26$\pm$1.99\% to 63.46$\pm$1.38\%, indicating their inability to adapt to diverse staining conditions. This suggests that purely visual models struggle to disentangle stain-induced variation from defocus, motivating the need for semantic conditioning. In contrast, FluoCLIP achieves 85.21$\pm$0.88\% (Table~\ref{tab:fluomix}), substantially outperforming all baselines.
The explicit integration of stain-specific text tokens with stain-conditioned rank prompts allows FluoCLIP to model complex stain-dependent focus variations that conventional FQA models cannot capture. FluoCLIP also maintains its advantage with a ViT-B/16 encoder (Table~\ref{tab:fluomix_vit}), demonstrating strong robustness across architectures. 

\noindent\textbf{Results on BBBC006.} BBBC006~\citep{ljosa2012annotated} contains fluorescence images of U2OS cells with limited stain diversity and clear cell structures. Conventional FQA models and lightweight CNNs such as FocusLiteNN~\citep{wang2020focuslitenn} fail to capture subtle focus variations. FluoCLIP achieves 93.05\% accuracy (Table~\ref{tab:bbbc}), surpassing OrdinalCLIP by 2.4\% while maintaining lower variance. These results indicate that even in relatively homogeneous fluorescence datasets, explicit stain conditioning improves focus estimation by increasing sensitivity to stain-specific optical cues.

\noindent\textbf{Results on FocusPath.} FocusPath~\citep{hosseini2019encoding} consists of bright-field H\&E slides where focus variations are relatively uniform and stain-invariant across focal planes. Under these homogeneous conditions, conventional FQA models already achieve near-optimal performance. Adding stain-conditioned prompts in FluoCLIP introduces unnecessary modeling capacity and slightly disrupts this smooth rank structure, resulting in a modest drop in accuracy (Table~\ref{tab:focuspath}). This behavior aligns with our design hypothesis that stain-aware modeling is most beneficial in fluorescence settings with strong stain-dependent variability, but offers limited or negative gains in stain-invariant bright-field datasets.   

\subsection{Ablation Study}

\noindent\textbf{Contribution of Individual Components.} 
Explicit stain alignment and stain-conditioned ranking are critical to FluoCLIP. We validate this on FluoMix using a ResNet-50 CLIP backbone. Table~\ref{tab:ablation} summarizes the results. The baseline (A), identical to OrdinalCLIP~\citep{li2022ordinalclip}, achieves 83.12\% accuracy. Adding a non-learnable stain token (B) slightly increases accuracy by 0.09\%, indicating that simply appending stain information is ineffective because CLIP has no prior knowledge of fluorescence-specific stain semantics. Even when the stain token is made learnable (C), accuracy drops more substantially (-1.74\%), showing that introducing trainable text parameter without explicitly aligning them to visual features can inject uninformative or noisy semantics that interfere with learning. 
Introducing learnable stain tokens through the stain-grounding stage (D) improves accuracy to 84.28\%, confirming the benefit of learning stain-visual correlations. Finally, incorporating the stain-guided ranking stage (E) raises accuracy to 85.21\%, a total gain of +2.09\% over the baseline. These results verify that both stages contribute synergistically—progressively enhancing the model’s ability to capture stain semantics and model stain-specific focus variations.

\noindent\textbf{Effects of Learning Stain Representations.} 
The ablations on BiomedCLIP~\citep{zhang2024biomedclip} confirm the same trend observed in Table~\ref{tab:ablation}\textemdash explicitly understanding stains and modeling stain-guided ranks significantly improve performance (Table~\ref{tab:biomedclip}).
Despite these gains, FluoCLIP with the standard CLIP ViT-B/16 backbone achieves a much higher accuracy of 90.75\%, surpassing all BiomedCLIP variants.
We attribute this gap to two limitations of BiomedCLIP: (1) text encoder (PubMedBERT~\cite{gu2021domain}), which lacks exposure to optical descriptors such as ``sharp", ``blur" or actual non-optimal images, and (2) image encoder, which was trained on mixed biomedical imagery rather than fluorescence microscopy.  
These findings highlight the importance of explicitly learning stain-grounded textual embeddings, rather than relying on domain-specific pretraining alone.

\begin{table}[t!]
\centering
\caption{Performance Evaluation of FluoCLIP using BiomedCLIP~\citep{zhang2024biomedclip} and OrdinalCLIP~\citep{li2022ordinalclip} as backbones. (A)--(C) use BiomedCLIP, while (D) corresponds to our main model based on CLIP ViT/B-16. Results are reported as mean $\pm$ standard deviation over five runs on FluoMix.}
\vspace{-0.8em}
\scriptsize\addtolength{\tabcolsep}{-4pt}
\begin{tabular}{llcccc}
\toprule
Type & Configuration  &  Accuracy (\%) & PLCC $\uparrow$ & SRCC $\uparrow$ & MAE $\downarrow$ \\
\midrule
(A) & BiomedCLIP & 83.99$\pm$0.25 & 0.990$\pm$0.0003 & 0.988$\pm$0.0003 & 0.163$\pm$0.0027 \\
(B) & (A) + $S^{\text{plain}}$ & 84.35$\pm$0.27 & 0.990$\pm$0.0002 & 0.9883$\pm$0.0002 & 0.159$\pm$0.0027 \\
(C) & (B) + $\tilde{R}^S$ & 86.29$\pm$0.51 & 0.991$\pm$0.0004 & 0.989$\pm$0.0003 & 0.140$\pm$0.0055 \\
\midrule
(D) & FluoCLIP    & 90.75$\pm$0.52 & 0.994$\pm$0.0005 & 0.992$\pm$0.0004 & 0.093$\pm$0.0055\\ 
\bottomrule
\end{tabular}
\label{tab:biomedclip}
\vspace{-0.9em}
\end{table}

\vspace{-0.5em}
\section{Conclusion}
\label{sec:conclusion}
We introduced \textbf{stain-aware ordinal focus quality assessment (FQA)} for fluorescence microscopy, demonstrating that focus behavior is inherently \emph{stain-dependent} rather than governed by a single global ordering.   
To address this, we proposed \textbf{FluoCLIP}, a two-stage vision–language framework that learns stain grounding and enables stain-conditioned ordinal reasoning. Together with our dataset \textbf{FluoMix}, this establishes a practical foundation for modeling focus under biologically diverse fluorescence conditions. 
Extensive experiments show that FluoCLIP consistently outperforms both conventional FQA methods and recent vision–language baselines, highlighting the importance of explicitly modeling stain-dependent focus behavior rather than relying on shared ordinal structures.
While FluoCLIP assumes predefined stain categories, future work will explore automatic stain discovery and more general conditioning mechanisms to extend stain-aware FQA to broader and less structured imaging settings.



\section*{Acknowledgment}
This research was supported by the Bio \& Medical Technology Development Program of the National Research Foundation (NRF) Korea (MSIT; RS-2022-NR068424), the G-LAMP Program of the NRF Korea (RS-2025-25442252), and the Korea Basic Science Institute (National Research Facilities and Equipment Center) funded by the Ministry of Education, Republic of Korea (RS-2024-00436263, RS-2025-02310437).
{
    \small
    \bibliographystyle{ieeenat_fullname}
    \bibliography{main}
}

\clearpage
\setcounter{page}{1}
\maketitlesupplementary

\section{Spatial Frequency}
In Section~\ref{sec:analysis}, we used the spatial frequency (SF)~\cite{Eskicioglu1995ImageQM} metric 
as a quantitative proxy for image sharpness to analyze stain-dependent focus behavior across datasets. 
For completeness, we provide the exact formulation of SF below.

Given an image $I \in \mathbb{R}^{M \times N}$, the row and column frequency components are defined as
\begin{equation}
\begin{aligned}
RF &= \sqrt{\frac{1}{(M-1)N}
\sum_{i=1}^{M-1}\sum_{j=1}^{N} (I(i+1,j)-I(i,j))^2}, \\[4pt]
CF &= \sqrt{\frac{1}{M(N-1)}
\sum_{i=1}^{M}\sum_{j=1}^{N-1} (I(i,j+1)-I(i,j))^2}, \\[4pt]
SF &= \sqrt{RF^2 + CF^2}.
\end{aligned}
\label{eq:sfmetric}
\end{equation}
Higher SF values indicate sharper, more in-focus images, while lower values correspond to defocus.

\begin{figure*}[t!]
\centering
\includegraphics[width=1\linewidth]{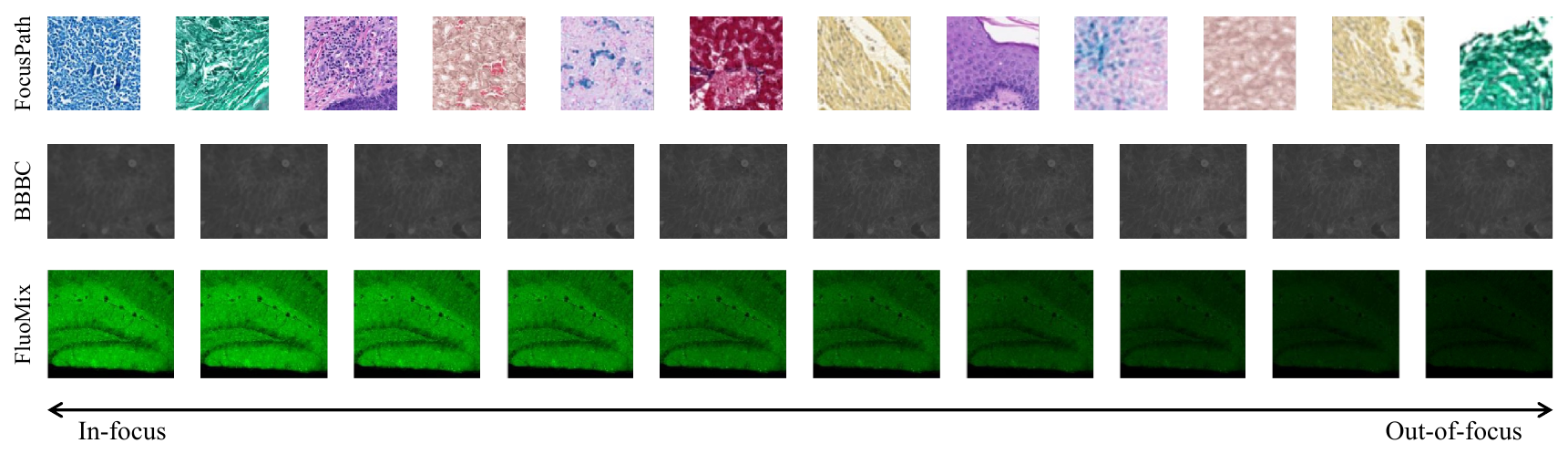}\\
\vspace{-0.8em}
\caption{Examples of dataset classes: The figure displays samples of images from three datasets across different focus levels. (Top) FocusPath dataset images numbered from 0 to 11, each showing different staining techniques. (Middle) BBBC006 dataset images numbered from 0 to 9.  (Bottom) FluoMix dataset images numbered from 0 to 9, representing different focus levels.}
\label{fig:datasets} 
\end{figure*}

\begin{table*}[t!]
\centering
\caption{
Overview of the FluoMix dataset, summarizing tissue types, fluorescent stains, 
and associated biological targets. Each stain is paired with distinct protein markers 
(e.g., neuronal, microglial, vascular, epithelial), reflecting the heterogeneity of 
fluorescence signals across tissues. The rightmost column reports the number of fields 
of view collected for each stain–target combination.
}
\label{tab:fluomix_overview}
\vspace{-0.3em}
\scriptsize
\addtolength{\tabcolsep}{13pt}

\begin{tabular}{lcccccc}
\toprule
\multicolumn{6}{c}{\textbf{Brain Tissue}} \\
\midrule

\textbf{Dataset} & \textbf{Hoechst 34580} & \textbf{Alexa 488} & \textbf{Cy3} & \textbf{Alexa 647} & \textbf{\# Sets} \\
\midrule
D1 & nucleus & Iba-1 & Tuj-1 & Collagen IV & 504 \\
D2 & nucleus & Neurofilament-M (NFM) & Tyrosine Hydroxylase (TH) & Collagen IV & 152 \\
D3 & nucleus & NeuN & Tyrosine Hydroxylase (TH) & Collagen IV & 554 \\
D4 & nucleus & GFAP & Tuj-1 & CD31 & 623 \\
\midrule
\addlinespace[4pt]

\toprule
\multicolumn{6}{c}{\textbf{Lung Tissue}} \\
\midrule
\textbf{Dataset} & \textbf{Hoechst 33342} & \textbf{Alexa 488} & \textbf{Cy3} & \textbf{Alexa 647} & \textbf{\# Sets} \\
\midrule
D5 & nucleus  & CD31 & Vimentin & Collagen IV & 634 \\
D6 & nucleus  & CD31 & Vimentin & Collagen IV & 596 \\
\midrule
\addlinespace[4pt]

\toprule
\multicolumn{6}{c}{\textbf{Liver Tissue}} \\
\midrule
\textbf{Dataset} & \textbf{DAPI} & \textbf{Alexa 488} & \textbf{Cy3} & \textbf{Alexa 647} & \textbf{\# Sets} \\
\midrule
D7 & nucleus  & CK19 & Claudin & ZO-1 & 96 \\
\midrule

\end{tabular}
\end{table*}

\begin{figure*}[t!]
\centering
\includegraphics[width=0.95\linewidth]{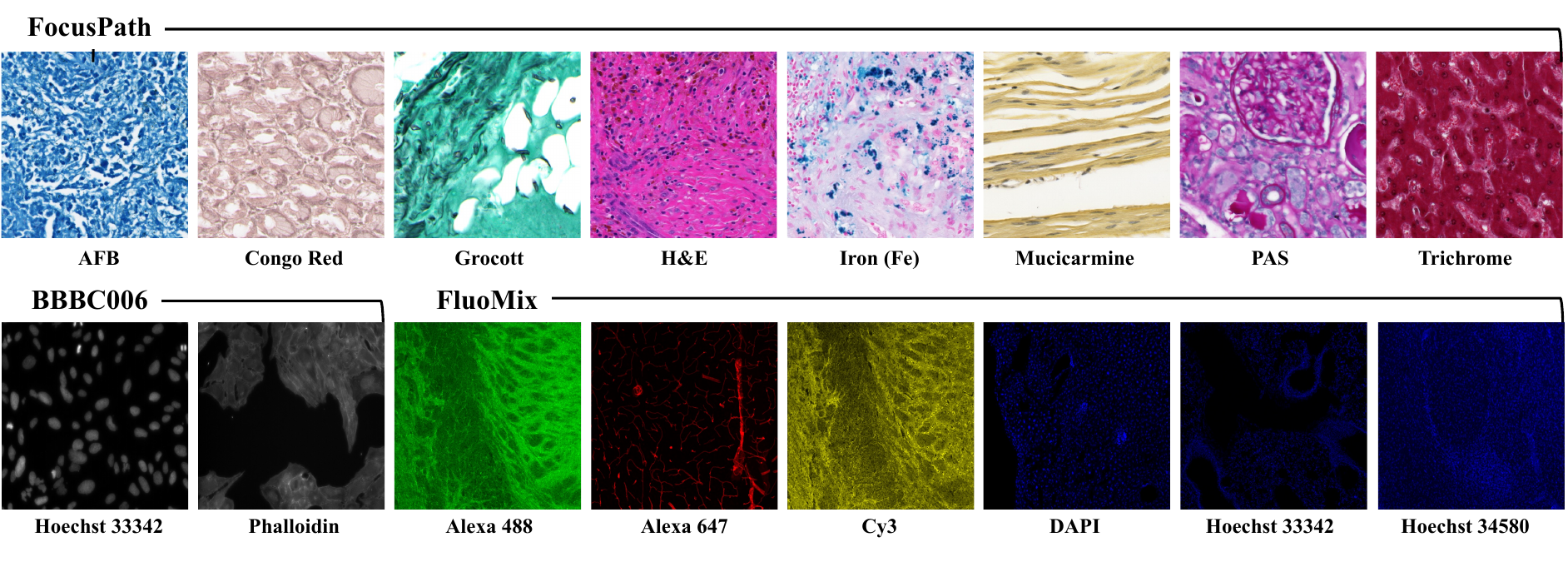}\\
\vspace{-0.8em}
\caption{Sample images illustrating stain diversity in the three datasets. (Top) \textbf{FocusPath}: bright-field microscopy images with eight different histological dyes. (Bottom left) \textbf{BBBC006}: fluorescence microscopy images of cell lines labeled with Hoechst 33342 and Phalloidin. (Bottom right) \textbf{FluoMix}: tissue-level fluorescence microscopy images covering six distinct fluorescent staining protocols.}
\label{fig:datasets_stain} 
\end{figure*}

\section{Dataset Details}
\subsection{Overview of Datasets}

We evaluated FluoCLIP on three datasets used in Section~\ref{sec:analysis} and Section~\ref{sec:experiments}: 
\textbf{FocusPath}~\citep{hosseini2019encoding}, \textbf{BBBC006}~\citep{ljosa2012annotated}, and our fluorescence microscopy dataset \textbf{FluoMix}. 
To ensure consistent training across datasets, all z-stacks were reorganized into discrete focus levels, ranging from in-focus to severely defocused, following the ordinal labeling protocol described in the main paper (Section~\ref{sec:dataset}). 
Representative examples from each dataset are shown in Figure~\ref{fig:datasets} and Figure~\ref{fig:datasets_stain}. The former illustrates focus-level (rank) variations across datasets, whereas the latter highlights stain-dependent visual differences across the three datasets.

\subsection{BBBC006}
BBBC006 dataset consists of fluorescence images of Hoechst 33342- and phalloidin-stained U2OS cells. 
Each field of view contains a 32-plane z-stack (2$\mu$m spacing), with the in-focus slice at position 16.  
Following prior work, we relabeled the stack into 10 ordinal focus levels (0–9).  
In our experiments, we use 23,341 images for training and 5,843 images for testing.

\subsection{FocusPath}
FocusPath provides 8,640 bright-field image patches (1024$\times$1024) from nine differently stained slides. 
Each patch is assigned an absolute z-level from 0 to 13. 
Due to limited samples in the most defocused slices (levels 12–13), we merge them into level 11, resulting in 12 ordinal levels (0–11).  
We follow the standard split of 3,876 training and 972 testing samples.

\subsection{FluoMix}
\label{sec:fluomix_details}
FluoMix is a fluorescence microscopy dataset designed to capture stain–dependent variability across multiple tissues. The dataset spans three tissue types (\textbf{brain}, \textbf{lung}, \textbf{liver}) and four fluorescent channels (Hoechst 33342/34580 or DAPI, Alexa Fluor 488, Cy3, Alexa Fluor 647). Each stain channel corresponds to distinct biological targets, such as neuronal markers (NeuN, NFM, TH), microglial markers (Iba-1), vascular markers (CD31, Collagen IV), and epithelial markers (CK19, Claudin, ZO-1). Table~\ref{tab:fluomix_overview} summarizes the stain–target combinations included in FluoMix.

Each field of view is acquired as a full z-stack (34 planes for FluoMix) covering the progression from in-focus to severely defocused slices. For consistency with FocusPath and BBBC006, all z-stacks are converted into \textbf{10 ordinal focus levels} (0 = in-focus, 9 = out-of-focus) using the relative labeling scheme described in Section~\ref{sec:dataset}. Representative examples are shown in Figure~\ref{fig:datasets}.

For the main experiments, we used brain-tissue subset of FluoMix (D1-D4), consisting of 25,967 training and 6,506 testing images across four stain combinations.

\section{Biological Preparation / Imaging Protocols}

FluoMix was constructed from adult \textbf{rat brain, lung, and liver} tissues, following institutional ethical 
guidelines approved by the Institutional Review Board (IRB). 
All samples were processed using standardized fixation, immunostaining, and confocal imaging protocols to 
ensure cross-tissue consistency while preserving stain-dependent optical characteristics.

\subsection{Tissue Preparation}
C57BL/6J male rats (8 weeks old) were euthanized using CO$_2$ anesthesia (30--70\% volume/min). 
Transcardial perfusion was performed with saline followed by fixation with 4\% paraformaldehyde (PFA).  
Brain, lung, and liver tissues were post-fixed for 24 hours, cryoprotected in 30\% sucrose, and sectioned into 
40~$\mu$m slices.

\subsection{Immunostaining}
Free-floating sections from all tissues were blocked in PBS containing 3\% bovine serum albumin (BSA) 
and 0.1\% Triton X-100 for 1~hour at room temperature.  
Slices were incubated overnight at 4$^\circ$C with primary antibodies targeting diverse cellular structures, including:

\begin{itemize}
    \item \textbf{Brain}: Iba-1, NeuN, NFM, Tuj-1, GFAP, TH, CD31, Collagen IV
    \item \textbf{Lung}: CD31, Vimentin, Collagen IV
    \item \textbf{Liver}: CK19, Claudin, ZO-1
\end{itemize}

After washing, sections were incubated with Alexa Fluor–conjugated secondary antibodies 
(Alexa Fluor 488, Cy3, Alexa Fluor 647).  
Hoechst 33342, Hoechst 34580 or DAPI (1:2000) was used to counterstain nuclei.  
Slides were washed, mounted with anti-fade medium, and allowed to dry under light-protected conditions.

\subsection{Imaging Conditions}
All fluorescence images were acquired using a Leica MICA confocal microscope with a 10$\times$ objective 
(NA 0.32) at a resolution of 1600~$\times$~1352 pixels (0.586~$\mu$m/pixel).  
Excitation wavelengths were selected based on fluorophore spectra:  
405~nm (Hoechst, DAPI), 488~nm (Alexa Fluor 488), 561~nm (Cy3), and 647~nm (Alexa Fluor 647), with matched bandpass emission filters to minimize spectral bleed-through.

To capture focus variability across staining conditions, each field of view (FOV) was acquired as a full z-stack:
\begin{itemize}
    \item \textbf{Brain}: 34-plane stacks (1-2~$\mu$m spacing)
    \item \textbf{Lung}: 34-plane stacks (2~$\mu$m spacing)
    \item \textbf{Liver}: 34-plane stacks (2~$\mu$m spacing)
\end{itemize}
Laser power and exposure settings were adjusted per channel to optimize signal-to-noise ratio while maintaining 
consistency across tissue types.  
No deconvolution or contrast enhancement was applied to preserve the native optical properties of each stain.

\begin{table*}[t]
\centering
\caption{Overview of the three few-shot evaluation settings used to assess label efficiency, cross-stain transfer, and cross-tissue generalization.}
\label{tab:fewshot_overview}
\scriptsize\addtolength{\tabcolsep}{0pt}
\vspace{-0.8em}
\begin{tabular}{p{2cm} p{2.3cm} p{3cm} p{3cm} p{5cm}}
\toprule
Generalization Type & Model Initialization & Adaptation Stain & Adaptation Tissue & Description \\
\midrule
\midrule
Few-Shot Learning
& CLIP-pretrained \newline on Natural Images
& Alexa 488, Cy3, Alexa 647, \newline Hoechst 34580 
& Brain 
& Test whether the model can learn stain-aware FQA with only k-shot samples  \\
\midrule
Cross-Tissue  
& FluoCLIP-pretrained \newline on Brain (D1-D4)
& Alexa 488, Cy3, Alexa 647\newline (Seen Stains)
& Lung (Unseen Tissue)  
& Adaptation to new tissue with distinct morphology while keeping stains unchanged \\
\midrule
Cross-Stain \newline \& Cross-Tissue  
& FluoCLIP-pretrained \newline on Brain (D1-D4)
& Hoechst 33342, DAPI\newline(Unseen Stains) 
& Lung, Liver (Unseen Tissues)
& Most challenging: test adaptation to entirely new stain-tissue pairs \\

\bottomrule
\end{tabular}
\end{table*}

\begin{table*}[t!]
\centering
\caption{Comparison of accuracy (\%) under few-shot learning settings on the FluoMix mouse brain dataset. CoOp follows the experimental protocol from the original paper~\cite{zhou2022learning}, using prompt tuning only. CoOp+$E_{img}$ extends CoOp by additionally fine-tuning the image encoder together with the prompts. OrdinalCLIP and FluoCLIP follow the main ResNet-50 experiment configuration described in Section~\ref{sec:experiments}. 
FluoCLIP exhibits the \textbf{steepest} improvement curve as supervision increases, demonstrating a strong stain-aware prior that enables rapid adaptation in low-data regimes. 
Results are reported as mean~$\pm$~standard deviation over five runs with different random seeds.}
\scriptsize\addtolength{\tabcolsep}{8pt}
\vspace{-0.8em}
\begin{tabular}{lcccccc}
\toprule
\#-Shots                &  1 & 8 & 16 & 32 & 64 & 128  \\
\midrule 
CoOp~\cite{zhou2022learning}
    & 11.72{\scriptsize$\pm$1.10}
    & 16.32{\scriptsize$\pm$0.91}
    & 17.88{\scriptsize$\pm$0.44}
    & 18.36{\scriptsize$\pm$0.61}
    & 18.38{\scriptsize$\pm$0.44}
    & 18.50{\scriptsize$\pm$0.50} \\

CoOp+$E_{img}$~\cite{zhou2022learning}
    & 13.58{\scriptsize$\pm$1.88}
    & 14.92{\scriptsize$\pm$1.10}
    & 14.98{\scriptsize$\pm$0.88}
    & 15.82{\scriptsize$\pm$1.17}
    & 17.06{\scriptsize$\pm$0.78}
    & 19.22{\scriptsize$\pm$1.17} \\

OrdinalCLIP~\cite{li2022ordinalclip}
    & 15.51{\scriptsize$\pm$1.15}
    & 19.43{\scriptsize$\pm$0.98}
    & 19.55{\scriptsize$\pm$0.29}
    & 20.60{\scriptsize$\pm$0.91}
    & 21.18{\scriptsize$\pm$0.50}
    & 19.99{\scriptsize$\pm$2.06} \\

FluoCLIP
    & 11.69{\scriptsize$\pm$1.15}
    & 15.92{\scriptsize$\pm$0.89}
    & 18.90{\scriptsize$\pm$1.32}
    & 20.45{\scriptsize$\pm$0.87}
    & 24.41{\scriptsize$\pm$0.52}
    & \textbf{29.11{\scriptsize$\pm$2.13}} \\
\bottomrule
\end{tabular}
\label{tab:fewshot_learn}
\end{table*}

\begin{table*}[t!]
\centering
\caption{Comparison of accuracy (\%) results under few-shot settings on a held-out fluorescence microscopy dataset acquired from \textbf{mouse lung tissue} with a \textbf{known staining combination} (Alexa 488, Cy3, and Alexa 647). We observe that from the 32-shot regime, FluoCLIP exhibits a clear advantage, suggesting that its stain-aware modeling begins to effectively leverage \textbf{stain-specific signal variation} when sufficient examples are available. The results are mean $\pm$ standard deviation over five random seeds.}
\scriptsize\addtolength{\tabcolsep}{8pt}
\vspace{-0.8em}
\begin{tabular}{lcccccc}
\toprule
\#-Shots  &  1 & 8 & 16 & 32 & 64 & 128  \\
\midrule 
OrdinalCLIP~\cite{li2022ordinalclip}
    & 20.09{\scriptsize$\pm$0.42}
    & 21.87{\scriptsize$\pm$1.23}
    & 23.99{\scriptsize$\pm$0.50}
    & 29.14{\scriptsize$\pm$0.66}
    & 41.26{\scriptsize$\pm$1.01}
    & 61.73{\scriptsize$\pm$0.62} \\

FluoCLIP
    & 20.35{\scriptsize$\pm$0.79}
    & 20.95{\scriptsize$\pm$1.05}
    & 23.51{\scriptsize$\pm$0.62}
    & 29.43{\scriptsize$\pm$0.63}
    & 43.27{\scriptsize$\pm$1.19}
    & 67.97{\scriptsize$\pm$1.31} \\
\bottomrule
\end{tabular}
\label{tab:fewshot_tissue}
\end{table*}

\begin{table*}[t!]
\centering
\caption{Comparison of accuracy (\%) under few-shot settings on held-out fluorescence microscopy datasets of unseen \textbf{mouse lung, liver} tissues with an unseen staining combination (Hoechst 33342 and DAPI). Because both stains target the nucleus and are visually similar to Hoechst 34580 used during base training, OrdinalCLIP performs better in the extremely low-shot regime. However, from the 32-shot setting onward, FluoCLIP shows a clear advantage, demonstrating its ability to leverage \textbf{stain-aware adaptation} once sufficient examples become available. The results are reported as mean $\pm$ standard deviation over five random seeds. }
\scriptsize\addtolength{\tabcolsep}{8pt}
\vspace{-0.8em}
\begin{tabular}{lcccccc}
\toprule
\#-Shots  &  1 & 8 & 16 & 32 & 64 & 128  \\
\midrule 
OrdinalCLIP~\cite{li2022ordinalclip}    
    & 19.82{\scriptsize$\pm$1.85}
    & 25.36{\scriptsize$\pm$1.23}
    & 29.37{\scriptsize$\pm$0.68}
    & 34.46{\scriptsize$\pm$1.25}
    & 42.82{\scriptsize$\pm$0.74}
    & 54.35{\scriptsize$\pm$1.17} \\
FluoCLIP
    & 18.75{\scriptsize$\pm$2.04}
    & 24.54{\scriptsize$\pm$0.93}
    & 29.20{\scriptsize$\pm$1.00}
    & 34.90{\scriptsize$\pm$1.45}
    & 44.57{\scriptsize$\pm$1.18}
    & 57.71{\scriptsize$\pm$0.46} \\
\bottomrule
\end{tabular}
\label{tab:fewshot_gen}
\end{table*}

\section{Few-Shot Validations}

Fluorescence microscopy exhibits substantial stain- and tissue-dependent variability, making it unrealistic to obtain dense focus annotations across all stain–tissue combinations encountered in practice.
As a result, few-shot adaptation is a key requirement for stain-aware FQA models deployed in real imaging workflows, where supervision is scarce and domain shifts naturally arise across laboratories.

To evaluate FluoCLIP under these realistic constraints, we consider three complementary few-shot settings:
(1) label-efficient learning with limited training samples,
(2) generalization to unseen tissues under known stains, and
(3) generalization to unseen stain–tissue pairs. 
As summarized in Table~\ref{tab:fewshot_overview}, these settings vary in whether the stain, the tissue, or both differ from the base training domain, thereby separating the effect of supervision scarcity from the effects of stain- and tissue-driven domain shifts.
This structured evaluation allows us to systematically assess how well FluoCLIP transfers its stain-grounded representations under increasingly challenging and practically relevant scenarios.

\subsection{Few-Shot Learning on FluoMix} 
\label{sec:fewshot_finetune}
To evaluate whether FluoCLIP can model Stain-Aware FQA with only a small number of labeled training samples, we first perform few-shot fine-tuning on the FluoMix brain-tissue subset (D1-D4). This restricted setup allows us to assess how effectively the model learns stain-dependent focus behavior when supervision is severely limited.
All methods were initialized from the same pretrained CLIP image encoder (ResNet50) and text encoders. For baselines, we finetuned CoOp~\cite{zhou2022learning} following the protocol from the original paper, where the encoders are frozen and only the prompt embeddings are learned. Because both OrdinalCLIP~\cite{li2022ordinalclip} and FluoCLIP update both the image encoder and text prompts, we additionally include CoOp+$E_{img}$, which extends CoOp by also fine-tuning the image encoder to ensure a fair architectural comparison. 
 
For each method, we sample $k$ labeled samples per focus rank per stain ($k=\{1, 8, 16, 32, 64, 128\}$). Since D1-D4 contain four stain combinations and 10 ordinal focus ranks, each few-shot training set consists of $4 \times 10 \times k$ training images. 
Although all models start from the generic CLIP checkpoint, this experiment evaluates whether they can adapt to the fluorescence FQA domain using only this small subset. In particular, it measures how effectively each method acquires stain-aware focus behavior when fine-tuning from a natural-image pretraining domain to the fluorescence microscopy domain with extremely limited supervision. This setting allows us to directly test whether FluoCLIP’s stain-aware design provides a stronger and more data-efficient adaptation capability than stain-agnostic baselines.

Adapting from CLIP to the fluorescence FQA domain with only a handful of labeled examples is inherently challenging, particularly because stain-dependent appearance variations amplify the effect of limited supervision. 
Nevertheless, the few-shot results in Table~\ref{tab:fewshot_learn} reveal consistent behavioral patterns across methods.  
CoOp performs the worst, and CoOp+$E_{img}$ often degrades further, consistent with observations from~\cite{zhou2022learning} that naive image-encoder finetuning can be harmful.  
OrdinalCLIP slightly outperforms FluoCLIP at very low shot counts (1–32), but its gains saturate quickly.  
In contrast, FluoCLIP exhibits the steepest improvement and ultimately achieves the best performance in the 64–128 shot range. 
Taken together, these results demonstrate that FluoCLIP’s stain-aware design provides a more data-efficient adaptation mechanism, enabling the model to acquire stain-dependent focus behavior more rapidly than stain-agnostic baselines once modest supervision is available.

\subsection{Few-Shot Generalization to Unseen Tissues}
\label{sec:fewshot_tissue}

Next, we evaluated FluoCLIP along the \emph{tissue generalization} axis by testing its ability to adapt to previously unseen tissues while keeping the stain conditions fixed. 
The model was pretrained on brain tissue (D1–D4) and then adapted using $k$ labeled samples per focus rank per stain from lung-tissue images (D5) stained with Alexa 488, Cy3, and Alexa 647.
For this setting, the few-shot training set is constructed using the same sampling protocol--$k$ labeled samples per focus rank for each stain ($k=\{1, 8, 16, 32, 64, 128\}$)--resulting in $3 \times 10 \times k$ training samples. 

Tissue-dependent variations introduce substantial domain shifts in cellular morphology, density, and spatial texture, making this setting more challenging. 
As shown in Table~\ref{tab:fewshot_tissue}, OrdinalCLIP~\cite{li2022ordinalclip} performs slightly better in the lowest-shot regime due to its stain-agnostic design, which avoids committing to stain priors that may behave differently across tissues.  However, FluoCLIP again shows a sharp performance increase as $k$ grows, surpassing OrdinalCLIP from the 32-shot regime onward.  
This indicates that FluoCLIP’s stain-grounded modeling generalizes effectively across tissue-dependent fluorescence variations once modest supervision becomes available.

\subsection{Few-Shot Generalization to Unseen Stain \& Tissue Pairs}
\label{sec:fewshot_gen}

We now evaluate the most challenging generalization setting, where both the stain and the tissue differ from those seen during base training. Starting from a checkpoint trained on the FluoMix brain-tissue subset (D1–D4), we adapt the model to two entirely unseen nuclear stains--Hoechst 33342 (D5–D6) and DAPI (D7)--each acquired from different tissues than the training domain. Adaptation uses only $k$ labeled samples per focus rank for each stain ($k=\{1, 8, 16, 32, 64, 128\}$), resulting in $2 \times 10 \times k$ training images. 
To ensure comparability with earlier experiments, we restrict evaluation to nucleus-targeting stains (DAPI, Hoechst 33342). Since these stains highlights the same biological structure, morphological differences caused by tissue variability are reduced, allowing stain-dependent optical differences to be examined more clearly (Figure~\ref{fig:datasets_stain})

As shown in Table~\ref{tab:fewshot_gen}, OrdinalCLIP~\cite{li2022ordinalclip} slightly outperforms FluoCLIP at 1–16 shots. Because all unseen stains in this setting target the nucleus and therefore appear visually similar to Hoechst 34580 stained images from the base training domain, OrdinalCLIP’s stain-agnostic prompts generalize reasonably well in this extremely low-shot regime.
In contrast, FluoCLIP has insufficient samples to learn the subtle stain-specific variations. However, as more labeled samples become available, FluoCLIP quickly surpasses all baselines and widens the performance gap at 32–128 shot range. This demonstrates that explicit stain grounding provides superior adaptability when both the stain and tissue shift simultaneously, becoming increasingly advantageous once even modest supervision is available.

\section{Ablation Study}
To understand the contribution of each component in FluoCLIP, we conduct a systematic ablation by varying the stain-token type, the use of the two-stage grounding pipeline, and the inclusion of the text-side adapter. 

We compare three types of stain tokens: $S^{plain}$ is a non-learnable text token that simply inserts the stain name, $S^{train}$ is a learnable token updated during training without explicit grounding, and $S$ is our grounded stain token obtained through the two-stage procedure. 
Table~\ref{tab:suppl_ablation} varies two factors that jointly define the stain-token type—Fine-tune ($S$) (trainable vs.\ frozen tokens) and Two Stage (with vs.\ without stain grounding)—along with an additional orthogonal choice of whether to include the text-side Adapter after CLIP text encoder.

With this ablation, we evaluate how different learning strategies for stain tokens-from frozen tokens to naïve trainable tokens to grounded tokens-and the presence of adapter affect the model's ability to acquire robust stain-aware focus representations.

\begin{table}[t!]
\centering
\caption{Ablation of FluoCLIP's learning strategy. We evaluate how stain tokens should be trained and whether the text-side adapter provides additional gains. Configuration (A-F) combines three stain-token variants ($S^{plain}$: frozen, $S^{train}$: naïvely trained, $S$: grounded via two-stage learning) with optional use of the grounding pipeline (Two Stages) and the adapter (Adapter). Results (mean $\pm$ std over five seeds) isolate the contribution of stain grounding, trainability, and adapter design.}
\scriptsize\addtolength{\tabcolsep}{-4pt}
\begin{tabular}{llcccc}
\toprule
Type & Configuration  & Fine-tune (S) & $\text{Two Stage}$ & $\text{Adapter}$ & Accuracy (\%)  $\uparrow$ \\
\midrule
\midrule
(A) & $S^{\text{plain}}$ &  $\times$ & $\times$ & $\times$ &81.04$\pm$0.81 \\
(B) & $S^{\text{plain}}$ + Adapter &  $\times$ & $\times$ & $\text{o}$ &83.21$\pm$3.93 \\
\midrule
(C) & $S^{\text{train}}$  & $\text{o}$ & $\times$ & $\times$ &80.50$\pm$0.67 \\
(D) & $S^{\text{train}}$ + $\text{Adapter}$ & $\text{o}$ & $\times$ & $\text{o}$ &81.38$\pm$0.63 \\
\midrule
(E) & $S$ & $\text{o}$ & $\text{o}$ & $\times$ & 81.83$\pm$0.95 \\
(F) & $S$ + $\text{Adapter}$ & $\text{o}$ & $\text{o}$ & $\text{o}$ & 84.28$\pm$0.88 \\
\bottomrule
\end{tabular}
\label{tab:suppl_ablation}
\end{table}

\subsection{Effects of Stain-Grounding} 
Configuration (A) uses non-learnable stain tokens ($S^{plain}$), while (C) replaces it with learnable stain tokens ($S^{train}$) but without the grounding stage (Two Stages). 
Interestingly, (C) performs worse than (A), indicating that simply making stain tokens trainable does not help. Without the grounding phase, the model receives no visual constraint linking stain-specific text features to fluorescence appearance, and learning tends to inject noise rather than meaningful stain semantics. 

When the grounding phase is enabled (E), the performance increases over both (A) and (C) (by +0.79\% and +1.33\%, respectively), demonstrating that explicit grounding is essential for producing stain tokens that actually reflect stain-dependent optical cues. Importantly, these gains appear even before introducing the adapter, confirming that grounding--not trainability alone--is the key factor.  
\subsection{Effects of Adapter} 
Configuration (B) and (D) evaluate the adapter with non-grounded stain tokens ($S^{plain}$ in (B), $S^{train}$ in (D)). The adapter yields moderate improvements in both cases, but the gains remain limited because the underlying stain tokens are not aligned with visual features.
Thus, architectural augmentation alone cannot produce meaningful stain-aware embeddings—the adapter benefits only materialize when the stain tokens themselves are grounded.

\subsection{Combined Effects} 
Configuration (E) and (F) corresponds to the two-stage FluoCLIP pipeline, with (F) additionally using the text-side adapter. 
Grounded stain tokens alone (E) already outperform all non-grounded alternatives (A, C), verifying that the grounding stage is the primary source of performance improvement. 
Adding the adapter (F) achieves the best results overall (84.28\%), confirming that grounded stain semantics and the adapter contribute complementary benefits. 
Overall, these ablations demonstrate that
(i) stain-grounding is necessary for producing stable and informative stain-aware embeddings, and
(ii) the adapter further refines these representations, but works best when grounding is present.
This explains why the full FluoCLIP pipeline achieves the strongest performance.

\begin{algorithm}[t!]
\caption{Overall Workflow of FluoCLIP}

\label{alg:total}
\begin{algorithmic}[1]
\Statex \hspace*{-1.5em}\textbf{Stage 1: Stain-Grounding}
\Statex \hspace*{-1.5em}\textbf{Input (Stage 1):}
\Statex Training set $\mathcal{D}_\text{stain}=\{(x_i, s_i)\}$
\Statex Stain tokens $\mathbf{S}$, adapter $\mathcal{A}$
\Statex Context tokens $C = \{[\textit{stain-context}]\}$
\Statex Image encoder $\mathrm{E_{img}}$
\Statex Text encoder $\mathrm{E_{text}}$

\Statex \hspace*{-1.5em}\textbf{Output (Stage 1):}
\Statex Updated $\mathbf{S}, \mathcal{A}$

\For{epoch $=1$ to \text{stage1}\_\text{max}\_\text{epoch}}
    \State Generate Prompt $\mathcal{P}^{\text{stain}}$ (Eq.~\ref{eq:prompt_stain})
    
    \For{ $(x_i, s_i) \in \mathcal{D}_\text{stain}$}
        \State Encode Image $v_i \gets \mathrm{E_{img}}(x_i)$
        
        \State Encode Text $t_{l}^{stain} \gets \mathcal{A}\big(\mathrm{E_{text}}(\mathcal{P}^{\text{stain}})\big)$
        
        \State Calculate Similarity $z^{stain}_{i,l} = \langle v_i, t_{l}^{stain} \rangle$
        
        \State Update $\mathbf{S}, \mathcal{A}$
        
    \EndFor
\EndFor

\Statex \vspace{0.5em}
\Statex \hspace*{-1.5em}\textbf{Stage 2: Stain-Guided Ranking}
\Statex \hspace*{-1.5em}\textbf{Input (Stage 2):}
\Statex Training set $\mathcal{D}_\text{rank}=\{(x_i, s_i, r_i)\}$
\Statex Grounded stain tokens $\mathbf{S}$, adapter $\mathcal{A}$ from Stage 1
\Statex Base ranks $\mathbf{R}^{\text{base}}$
\Statex Conditioning network $f_\theta$
\Statex Context tokens $C = \{[\textit{stain-context}], [\textit{rank-context}]\}$
\Statex \hspace*{-1.5em}\textbf{Output (Stage 2):}
\Statex Updated $\mathbf{R}^{\text{base}}$, $f_\theta$, $\mathrm{E_{img}}, C$

\For{epoch $=1$ to \text{stage2}\_\text{max}\_\text{epoch}}
    \State Generate Stain-Guided Rank $\tilde{\mathbf{R}}_{k}^{l}$ (Eq.~\ref{eq:2nd_feat},~\ref{eq:rank_tilde})
    \For{$(x_i, s_i, r_i) \in \mathcal{D}_\text{rank}$}
        \State Encode Image $v_i \gets \mathrm{E_{img}}(x_i)$
        \State Generate Prompt $\mathcal{P}^{\text{rank},i}$ (Eq.~\ref{eq:prompt_rank})
        \State Encode Text $t^{\text{rank},i} \gets \mathcal{A}\big(\mathrm{E_{text}}(\mathcal{P}^{\text{rank},i})\big)$
        
        \State Calculate Similarity $z^{rank, i}_{k} = \langle v_i, t^{\text{rank},i}_k \rangle$ 
        
        \State Update $\mathbf{R}^{\text{base}}, f_\theta, \mathrm{E_{img}}, C$ 
    \EndFor
\EndFor
\end{algorithmic}
\end{algorithm}

\section{Algorithm}
Algorithm~\ref{alg:total} outlines the overall two-stage pipeline of FluoCLIP. \textbf{Stage~1} performs \textit{stain-grounding}, in which the stain tokens and adapter are optimized with stain-class supervision to align vision–language representations with fluorescence-specific stain semantics.
\textbf{Stage~2} conducts \textit{stain-guided ranking}. The grounded stain embeddings modulate the base rank embeddings through a conditioning network, and interpolation yields continuous stain-conditioned rank tokens. These rank tokens, together with learnable context tokens and the fixed grounded stain tokens, are assembled into sample-specific sentences and encoded by the frozen CLIP text encoder. Image–text similarities are then used to predict the focus rank, while the base rank tokens, context tokens, conditioning network, and image encoder are updated with ranking supervision.

\section{Biomedical CLIP Variants}

To evaluate whether domain-specific pretraining alone suffices for fluorescence FQA, we compare FluoCLIP with biomedical CLIP variants, including PubMedCLIP~\citep{eslami2023pubmedclip} and PMC-CLIP~\citep{lin2023pmc}. These models are pretrained on large-scale biomedical image–text corpora and provide stronger domain alignment than the original CLIP.

As shown in Table~\ref{tab:biomedicalclip}, both PubMedCLIP and PMC-CLIP achieve competitive performance on FluoMix, outperforming standard image-only baselines. In particular, PMC-CLIP benefits from broader biomedical coverage and achieves higher accuracy than PubMedCLIP.

However, FluoCLIP consistently outperforms both models, achieving the highest accuracy. This result indicates that domain-specific pretraining alone is insufficient for FQA, as it does not explicitly capture stain-dependent optical variations or ordinal focus structure. By incorporating stain-aware grounding and stain-conditioned ordinal modeling, FluoCLIP effectively resolves these limitations and yields more consistent and accurate focus predictions across diverse fluorescence conditions.

\begin{table}[t]
\centering
\caption{Comparison with biomedical CLIP variants on FluoMix. Each method uses its own pretrained ResNet50-based vision encoder corresponding to its original pretraining. Results are reported as mean $\pm$ standard deviation over five runs.}
\scriptsize
\begin{tabular}{lccc}
\toprule
Method & PubMedCLIP & PMC-CLIP & FluoCLIP \\
\midrule
Accuracy (\%) & 81.48$\pm$0.34 & 84.06$\pm$0.23 & 85.21$\pm$0.88 \\
\bottomrule
\end{tabular}
\label{tab:biomedicalclip}
\vspace{-0.5em}
\end{table}

\clearpage

\end{document}